\documentclass[12pt]{iopart}

\usepackage{iopams}  
\usepackage{graphicx}
\usepackage{xcolor}
\usepackage{hyperref}
\usepackage{subfigure}
\usepackage{cite}
\expandafter\let\csname equation*\endcsname\relax
\expandafter\let\csname endequation*\endcsname\relax
\usepackage{amsmath}

\usepackage{cleveref}

\begin{document}

\title{Graph Neural Networks in Particle Physics}

\author{Jonathan Shlomi$^1$, Peter Battaglia$^2$, Jean-Roch Vlimant$^3$}

\address{$^1$ Weizmann Institute of Science, Rehovot, Israel}
\address{$^2$ DeepMind, London UK}
\address{$^3$ California Institute of Technology, PMA, Pasadena, CA, USA 91125-0002}

\ead{jvlimant@caltech.edu}
\vspace{10pt}
\begin{indented}
\item[]July 2020
\end{indented}

\begin{abstract}
Particle physics is a branch of science aiming at discovering the fundamental laws of matter and forces.
Graph neural networks are trainable functions which operate on graphs---sets of elements and their pairwise relations---and are a central method within the broader field of geometric deep learning. They are very expressive and have demonstrated superior performance to other classical deep learning approaches in a variety of domains.
The data in particle physics are often represented by sets and graphs and as such, graph neural networks offer key advantages.
Here we review various applications of graph neural networks in particle physics, including different graph constructions, model architectures and learning objectives, as well as key open problems in particle physics for which graph neural networks are promising.
\end{abstract}

%
%
%
%
%


\section{Introduction}
\label{sec:intro}
Particle physics focuses on understanding fundamental laws of nature by observing elementary particles, either in controlled environments (collider physics) or in nature (astro-particle).
The standard model of particle physics is a theory of the strong, weak and electromagnetic forces, and elementary particles (quarks and leptons).
Physicists are building experiments to measure elementary particles and by using statistical methods can test the validity of various models.
The data from the experiments are generally a sparse sampling of a physics process in both time and space.

Machine learning has historically played a significant role in particle physics \cite{Radovic:2018dip}, with classification and regression applications using classical techniques, such as boosted decision trees, support vector machine, simple multi-layer perceptrons, etc.
Inspired by the success deep learning has achieved at reaching super-human performance at various tasks, various domains in the physical sciences~\cite{Carleo:2019ptp}, including particle physics~\cite{Radovic:2018dip,Guest:2018yhq,Bourilkov:2019yoi,Larkoski:2017jix,livingreview}, have begun exploring deep learning as a unique tool for handling difficult scientific problems that go beyond straightforward classification, to organize and make sense of vast data sources, draw inferences about unobserved causal factors, and even discover physical principles underpinning complex phenomena~\cite{cranmer2019learning,cranmer2020discovering}.

\begin{figure}[h]
	\centering
	\includegraphics[]{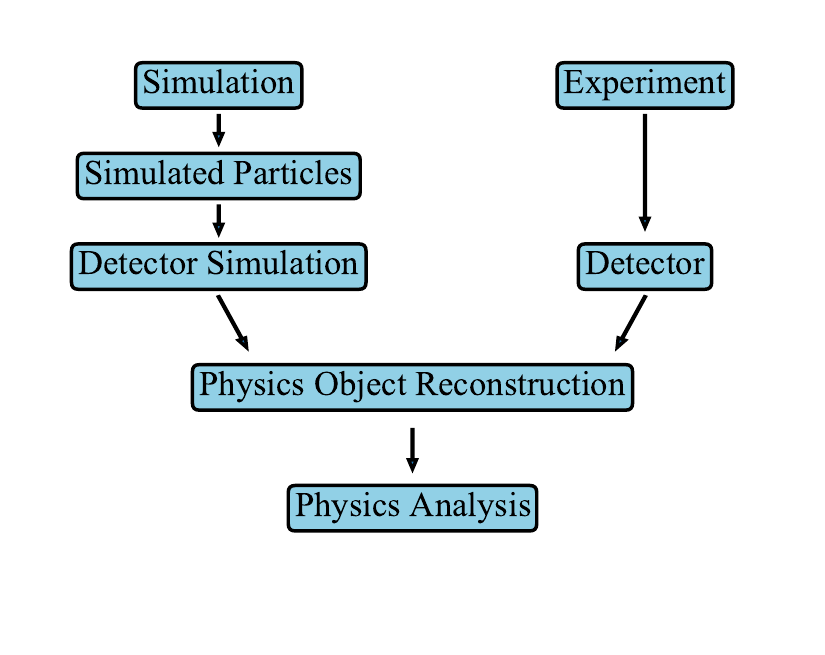}
	\caption{Simulation is used in HEP experiments to create a "truth record" of the physics event which caused a certain detector response. This "truth record" is used to train supervised learning algorithms to invert the detector simulation and infer something about the underlying physics from the observed data. These algorithms are then applied to real data that were measured by the detector.}
	\label{fig:simulation_chain}
\end{figure}

HEP experiments often use machine learning for learning complicated inverse functions, trying to infer something about the underlying physics process from the information measured in the detector. This scheme is illustrated in figure~\ref{fig:simulation_chain}.

While the most widely used trio of deep learning building blocks---the fully connected network (FC), convolutional neural network (CNN) and recurrent neural network (RNN)---have proven valuable across many scientific domains, the focus of this review is on a class of architectures called Graph Neural Networks (GNN) --- as described below, we regard self-attention as a graph-based architecture---, which can be trained from data to learn functions on graphs. 
Many problems involve data represented as unordered sets of elements with rich relations and interactions with one another, and can be naturally expressed as graphs. They are however not convenient to represent as vectors, grids, or sequences --- the format required by FCs, CNNs, and RNNs, respectively --- unless for specific structure of tree~\cite{mou2014convolutional,shen2018ordered}.
Extensive reviews of GNNs are available in the literature \cite{bronstein2017geometric,gilmer2017neural,battaglia2018relational,zhou2018graph,wu2019comprehensive}.
However applications of GNNs in high energy physics (HEP) are evolving rapidly, and the purposes of this review are to outline the key principles and uses of GNNs for particle physics, and build bridges between physics and machine learning by exposing researchers on both sides to important, challenging problems in each others' domains.

\paragraph{Data Representation.}\label{sec:representation}

Measurements in particle physics are commonly done in large accelerator facilities (CERN, KEK, Fermilab, etc), using detectors with sizes on the order of tens of meters, which capture millions of high-dimensional measurements each second.
These detectors are composed of multiple sub-detectors --- tracking detector, calorimeters, muon detector, etc --- each using a different technology to measure the trace of particles.
The data in particle physics are therefore heterogeneous.
Detectors in astrophysics are typically bigger, with size up to kilometers (IceCube, Antares, etc) constructed around a single measurement technology, the data are therefore homogeneous.
In both cases, the measurements are inherently sparse in space, due to the design of the geometry of the sensors.
The measurements therefore do not a-priori fit homogeneous, grid-like data structures. 

Deep learning is often applied on high level features derived from particle physics data~\cite{Radovic:2018dip}. This can improve over more classical data analysis methods, but does not use the full potential of deep learning, which can be effective when operating on lower level information.

\begin{figure}[h]
	\begin{center}
		\subfigure[]{
			\includegraphics[]{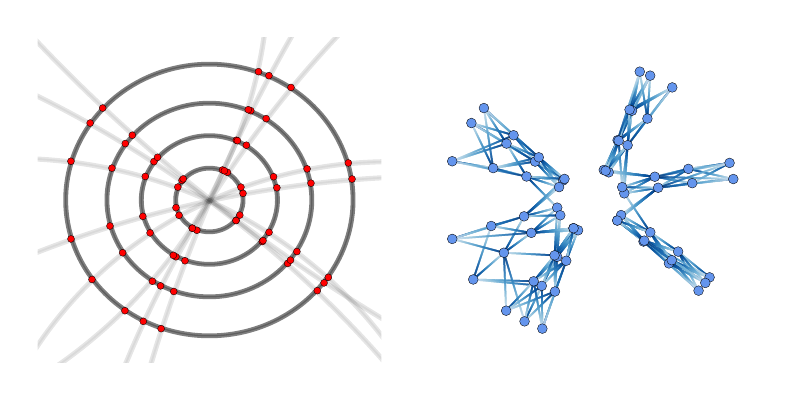}
		}
		\subfigure[]{
			\includegraphics[]{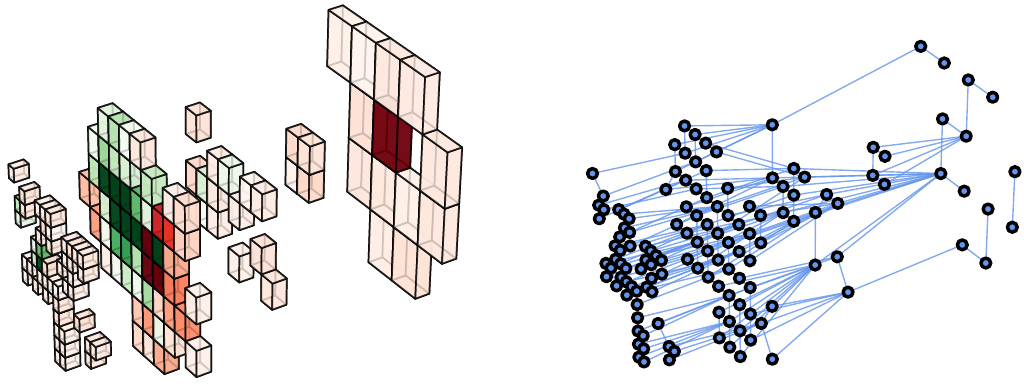}
		}
		\subfigure[]{
			\includegraphics[]{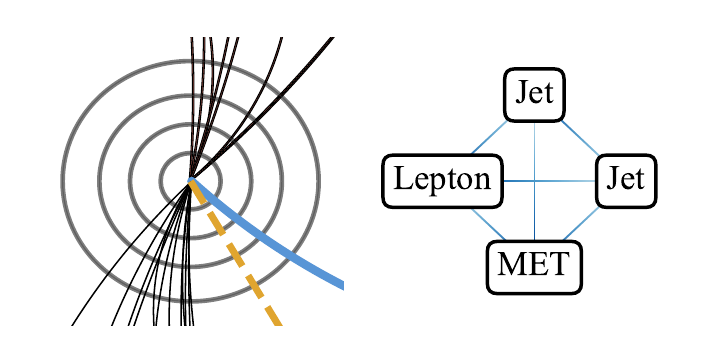}
		}
			\subfigure[]{
		\includegraphics[]{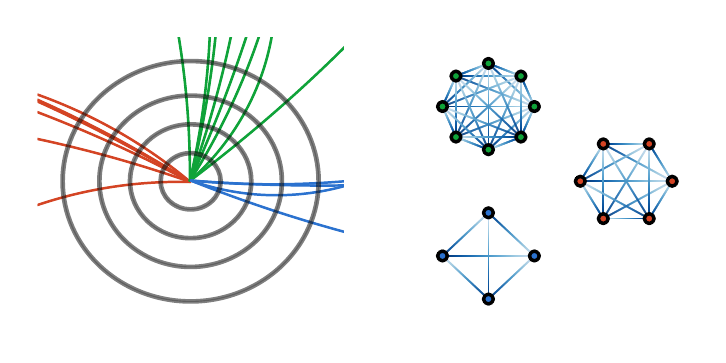}
	}

\caption{HEP data lend itself to being represented as a graph for many applications: (a) clustering tracking detector hits into tracks, (b) segmenting calorimeter cells, (c) classifying events with multiple types of physics objects, (d) jet classification based on the particles associated to the jet.}
		\label{fig:data_as_graph}
	\end{center}
\end{figure}

Some data in particle physics can be fractionally interpreted as images and hence computer vision techniques (CNNs) are being applied with improved performances \cite{ATLAS:2017dfg,Kasieczka:2017nvn,Macaluso:2018tck,Andrews:2018nwy,Lin:2018cin,ATLAS:2019fxb}.
However, image representations face some limitations with irregular geometry of detectors or sparsity of the projections applied. 
Because of the inherent loss of information, image representations may constrain the amount of information that can be extracted from the data.

Measurement and reconstructed objects can be viewed as sequences, with an order imposed from theoretical or experimental understanding of the data. 
Methods otherwise applied to natural language processing (e.g., RNNs, LSTMs~\cite{hochreiter1997long}, GRUs~\cite{cho2014learning}, etc) have thus been explored \cite{ATLAS:2017gpy,Sirunyan:2020lcu}.
While the ordering used can usually be justified experimentally, it is often imposed and therefore constrains how the data are presented to models. 
This ordering can also be learned \cite{Louppe:2017ipp} in some cases, using prior experimental knowledge of the physics process at stake.
This is however not always the case and one may expect that the imposed ordering will reduce the learning performance --- ordering that is not required as we will see in the following. 
For example~\cite{DIPs} shows evidence that a permutation invariant network outperforms a sequence based algorithm that uses the exact same input features, for the same classification task.

At many levels the data are, by definition, sets (unordered collection) of items.
If one considers relation between items (geometrical, or physical) a set transforms into a graph with the addition of an adjacency matrix.
There is a-priori less limitation in applying deep learning on this intrinsic representation of the data, than at the other levels mentioned above. 
A variety of HEP data and their formulation as graphs is illustrated in figure~\ref{fig:data_as_graph}.

We concentrate in this review on the applications of GNNs to HEP.
We argue why graphs are a very useful data representation, and review key architectures. 
Common traits in graph construction and model architecture will be linked to the specific requirements of the HEP problems under consideration. 
By providing a normalized description of the models through the formalism introduced in \cite{battaglia2018relational} we hope to make the adoption and further development of GNNs for HEP simpler.

This review paper is organized as follows. 
An overview of the field of geometrical deep learning is given in section~\ref{sec:geomdeeplearning}. Existing applications to particle physics are reviewed in \ref{sec:applications}.
General guidelines for formulating HEP tasks for GNNs are given in section~\ref{sec:guidelines}.
In particular we go in the details of the different approaches in building the graph connectivity in section~\ref{sec:graphconstruction}, the various model architecture adopted in section~\ref{sec:modelarch}.
This paper concludes with a discussion on the various approaches and the remaining open questions in section~\ref{sec:discussion}.

\section{Geometric Deep Learning}\label{sec:geomdeeplearning}
\subsection{Overview}

Deep learning has been central to the past decade's advances in machine learning and artificial intelligence~\cite{schmidhuber2015deep,lecun2015deep}, and can be understood as the confluence of several key factors.
First, large neural networks can express very complex functions. Second, valuable information in big data can be encoded into the parameters of large neural networks via gradient-based training procedures. 
Third, parallel computer hardware can perform such training in hours or days, which is efficient enough for many important use cases. Fourth, well-designed software frameworks, such as TensorFlow \cite{tensorflow2015-whitepaper} and PyTorch \cite{paszke2017automatic}, lower the technical bar to developing and distributing deep learning applications, making powerful machine learning tools broadly accessible to practitioners.

Fully connected, convolutional, and recurrent \textit{layers} have been the primary building blocks in modern deep learning,
each of which carries different \textit{inductive biases}, which incentivize or constrain the learning algorithm to prioritize one solution over another. 
For example, convolutional layers share their underlying kernel function across spatial dimensions of the input signal, while recurrent layers share across the temporal dimension of the input. These building blocks are most suitable for approximating functions on vectors, grids, and sequences, but when a problem involves data with richer structure, these modules are not always convenient or effective to apply. 
For example, consider learning functions over sets of particles --- while it is possible to order them, for example sorting by the transverse momentum $p_{T}$ of the particle, the imposed ordering in not unique,  and it fails to reflect that particles are fundamentally unordered. The aforementioned deep learning modules do not have appropriate inductive biases to exploit this richer graphical structure. 

Graph-structured data are ubiquitous across science, engineering, and many other problem domains. A graph is defined, minimally, as a set of nodes as well as a set of edges adjacent to pairs of nodes. 
Richer varieties and special cases include: trees, where there is exactly one sequence of edges connecting any two nodes; directed graphs, where the two nodes associated with an edge are ordered; attributed graphs, which include node-level, edge-level, or graph-level attributes; multigraphs, where more than one edge may exist between a pair of nodes; hypergraphs, where more than two nodes are associated with an edge; etc. 
Crucially, graphs are a natural and powerful way of representing many complex systems \cite{bronstein2017geometric,gilmer2017neural,battaglia2018relational,zhou2018graph,wu2019comprehensive}, e.g., trees for representing evolution of species, or the hierarchical structure of sentences; lattices and meshes for representing regular and irregular discretizations of space, respectively; dynamic networks for representing traffic on roads and social relationships over time.

GNNs~\cite{scarselli2008graph,bronstein2017geometric,gilmer2017neural,battaglia2018relational} are a class of deep learning architectures which implement strong relational inductive biases for learning functions that operate on graphs. 
They implement a form of parameterized message-passing whereby information is propagated across the graph, allowing sophisticated edge-, node-, and graph-level outputs to be computed. 
Within a GNN there are one or more standard neural network building blocks, typically fully connected layers, which implement the message computations and propagation functions. 
The first GNNs~\cite{gori2005new,scarselli2008graph} were developed and applied for network analysis, especially on internet data, and were trained not with the back-propagation algorithm, but with fixed point iteration via the Almeida-Pineda algorithm~\cite{almeida1987learning,pineda1987generalization}. 
Li et al.'s~\cite{li2015gated}'s gated graph sequence neural networks helped integrate more recent deep learning innovations into GNNs, adding RNN modules for improving multiple rounds of message-passing and optimizing their parameters by the back-propagation learning rule~\cite{schmidhuber2015deep,lecun2015deep}. 

In recent years, the field of GNNs has grown very rapidly, with applications to science and engineering. For example, graph convolution has been used for molecular fingerprinting~\cite{kearnes2016molecular}. Message-passing neural networks~\cite{gilmer2017neural}, which provided a general formulation of GNNs which captured a number of previous methods, were introduced for quantum chemistry. Interaction networks~\cite{battaglia2016interaction} and graph networks~\cite{battaglia2018relational} have been developed for learning to simulate increasingly complex physical systems~\cite{battaglia2016interaction,sanchez2018graph,li2018dpi,sanchez2020learning}. 

GNNs are situated within the broader family of what Bronstein et al.~\cite{bronstein2017geometric} term \textit{geometric deep learning}, which, aside from GNNs, captures related deep learning methods which apply to data structures beyond vectors, tensors, sequences, etc. 
Their survey explores graph signal processing and how it can be connected to deep learning, with substantial discussion on how the general principles of CNNs applied to Euclidean signals can be transferred to graph-structured signals. Key examples of spectral graph convolution approaches are~\cite{bruna2013spectral,defferrard2016convolutional,henaff2015deep}, which applied neural networks to the eigenvalues and eigenvectors of the graph Laplacian.

Much work on GNs has focused on learning physical simulation \cite{battaglia2016interaction,sanchez2018graph,li2018dpi,Ummenhofer2020Lagrangian}, similar to Lagrangian methods for particle-based simulation in engineering and graphics. The system is represented as a set of particle vertices, whose interactions are represented by edges and computed via learned functions. Recent work by \cite{sanchez2020learning} highlights how far this sub-field has advanced: they trained models to predict systems of thousands of particles, which represent fluids, solids, sand, and ``goop'', and show generalization to orders of magnitude more particles and longer trajectories than experienced during training. Because GNs are highly parallelizable on modern deep learning hardware (GPUs, TPUs, FPGAs), their approach scaled well, and its speed was on par with heavily engineered state-of-the-art fluid simulation engines, despite that they did not optimize for speed in their work.

Recently GNs have been extended by adding inductive biases derived from physics, adjusting their architectures to be consistent with Hamiltonian~\cite{sanchez2019hamiltonian} and Lagrangian mechanics~\cite{cranmer2020lagrangian}, which can improve performance and generalization on various physical prediction problems. Other recent work~\cite{cranmer2019learning} has shown symbolic physical laws can be extracted from the learned functions within a GN.

\subsection{The Graph Network Formalism }
\label{sec:GNformalism}
Here we focus on the \textit{graph network} (GN) formalism~\cite{battaglia2018relational}, which generalizes various GNNs, as well as other methods (e.g., Transformer-style self-attention~\cite{vaswani2017attention}). GNs are graph-to-graph functions, whose output graphs have the same node and edge structure as the input. Adopting \cite{battaglia2018relational}'s formalism, a graph can be represented by, $G = (\mathbf{u}, V, E)$, with $N_v$ vertices and $N_e$ edges. The $\mathbf{u}$ represents graph-level attributes. The set of \textit{nodes} (or vertices) are $V = \{\mathbf{v}_i\}_{i=1:N_v}$, where $\mathbf{v}_i$ represents the $i$-th node's attributes. The set of edges are $E = \{\left(\mathbf{e}_k, r_k, s_k\right)\}_{k=1:N_e}$, where $\mathbf{e}_k$ represents the $k$-th edge's attributes, and $r_k$ and $s_k$ are the indices of the two ($r$eceiver and $s$ender, respectively) nodes connected by the $k$-th edge.

A GN's stages of processing are as follows.
\begin{align}
  \begin{split}
    \mathbf{e}'_k &= \phi^e\left(\mathbf{e}_k, \mathbf{v}_{r_k}, \mathbf{v}_{s_k}, \mathbf{u} \right) \\
    \mathbf{v}'_i &= \phi^v\left(\mathbf{\bar{e}}'_i, \mathbf{v}_i, \mathbf{u}\right) \\
    \mathbf{u}' &= \phi^u\left(\mathbf{\bar{e}}', \mathbf{\bar{v}}', \mathbf{u}\right)
  \end{split}
  \begin{split}
    \mathbf{\bar{e}}'_i &= \rho^{e \rightarrow v}\left(E'_i\right) \hspace{48pt} \triangleright \text{ Edge block} \\
    \mathbf{\bar{e}}' &= \rho^{e \rightarrow u}\left(E'\right) \hspace{48pt} \triangleright \text{ Vertex block}\\
    \mathbf{\bar{v}}' &= \rho^{v \rightarrow u}\left(V'\right) \hspace{48pt} \triangleright \text{ Global block}
  \end{split}
  \label{eq:gn-functions}
\end{align}

\begin{figure}[]
	\centering
		   \subfigure[]{
	\includegraphics[]{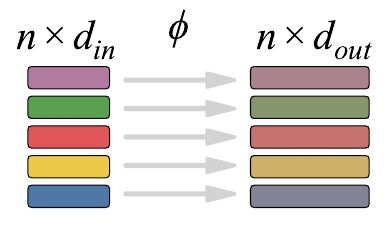}
	}
		   \subfigure[]{
	\includegraphics[]{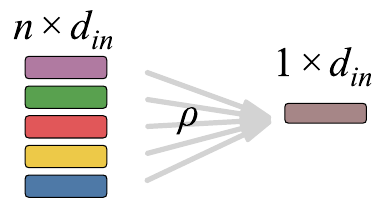}
		   }
	\caption{The internal components of a GN block are \textit{update functions} and \textit{aggregation functions}. (a) The update functions take a set of objects with a fixed size representation, and apply the same function to each of the elements in the set, resulting in an updated representation (also with a fixed size). (b) The aggregation functions take a set of objects and create one fixed size representation for the entire set, by using some order invariant function to group together the representations of the objects (such as an element-wise sum).}
	\label{fig:gn_functions}
\end{figure}

A GN block contains 6 internal functions: 3 \textit{update functions} ($\phi^e$, $\phi^v$, and $\phi^u$) and 3 \textit{aggregation functions} ($\rho^{e \rightarrow v}$, $\rho^{e \rightarrow u}$, and $\rho^{v \rightarrow u}$). The GN formalism is not a specific model architecture, it does not determine what exactly those functions are. The update functions are functions of fixed size input and fixed size output, and the aggregation functions take in a variable-sized set of inputs (such as a set of edges connected to a particular node) and output a fixed size representation of the input set. This is illustrated in figure~\ref{fig:gn_functions}.

\begin{figure}[thbt]
	\begin{center}
		\subfigure[]{
			\includegraphics[width=0.775\linewidth]{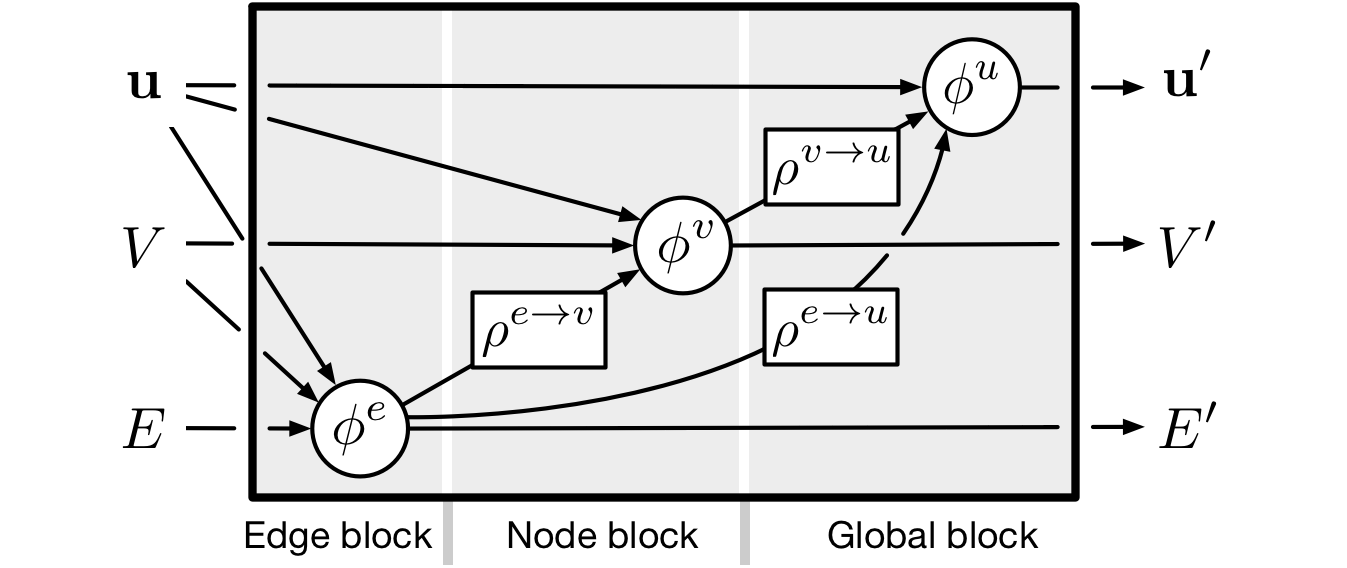}
			\label{fig:full_gn_block}
		}
	   \subfigure[]{
			\includegraphics[width=0.575\linewidth]{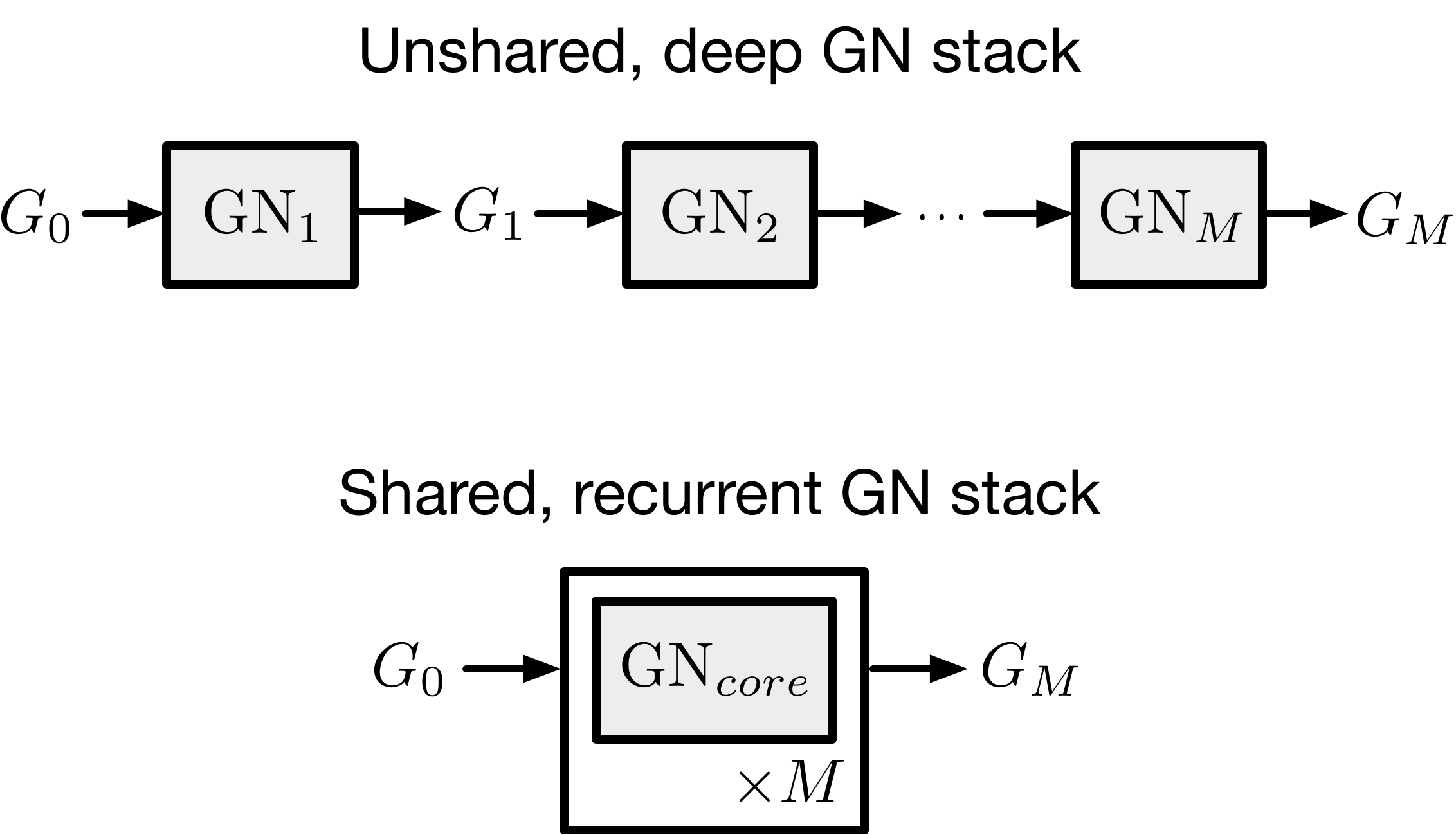}
			\label{fig:gn_stack}
		}
	
		\caption{(a) A GN block (from~\cite{battaglia2018relational}). An input graph, $G=(\mathbf{u}, V, E)$, is processed and a graph with the same edge structure but different attributes, $G'=(\mathbf{u}', V', E')$, is returned as output. The component functions are described in Equation~\ref{eq:gn-functions}. (b) GN blocks can be composed into more complex computational architectures. The top row shows a sequence of different GN blocks arranged in series, or depth-wise, fashion. The bottom row replaces the distinct GN blocks with a shared, recurrent, configuration.}
		\label{fig:gn-framework}

	\end{center}
\end{figure}

The \textit{edge block} computes one output for each edge, $\mathbf{e}'_k$, and aggregates them by their corresponding receiving node, $\mathbf{\bar{e}}'_i$, where $E'_i$ is the set of edges incident on the $i$-th node. The \textit{vertex block} computes one output for each node, $\mathbf{v}'_i$. The edge- and node-level outputs are all aggregated in order to compute the \textit{global block}. The output of the GN is the set of all edge-, node-, and graph-level outputs, $G'=(\mathbf{u}', V', E')$. See Figure~\ref{fig:gn-framework}a.

In practice the $\phi^e$, $\phi^v$, and $\phi^u$ are often implemented as a simple trainable neural network, e.g. a fully connected network. The $\rho^{e \rightarrow v}$, $\rho^{e \rightarrow u}$, and $\rho^{v \rightarrow u}$ functions are typically implemented as permutation invariant reduction operators, such as element-wise sums, means, or maximums. 
The $\rho$ functions must be permutation invariant if the GN block is to maintain permutation equivariance.

Some key benefits of GNs are that they are generic: if a problem can be expressed as requiring a graph to be mapped to another graph or some summary output, GNs are often suitable. They also tend to generalize well to graphs not experienced during training, because the learning is focused on the edge- and node-level---in fact if the global block is omitted, the GN is not even aware of the full graph in any of its computations, as the edge and node blocks take only their respective localities as input. Yet when multiple GN blocks are arranged in deep or recurrent configurations, as in Figure~\ref{fig:gn-framework}b, information can be processed and propagated across the graph's structure, to allow more complex, long-range computations to be performed.

\begin{figure}[!b]
	\centering
		   \subfigure[]{
	\includegraphics[height=4.0cm]{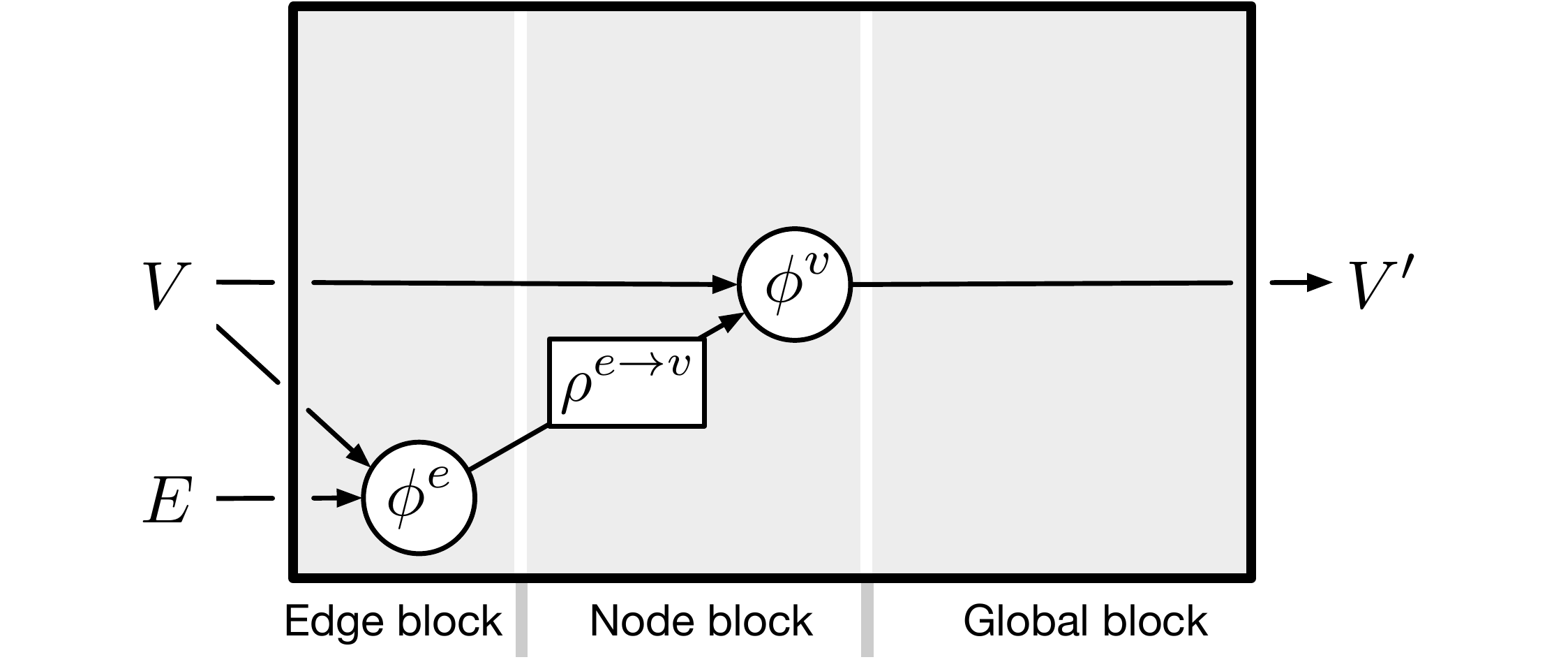}
	}
		   \subfigure[]{
	\includegraphics[height=4.0cm]{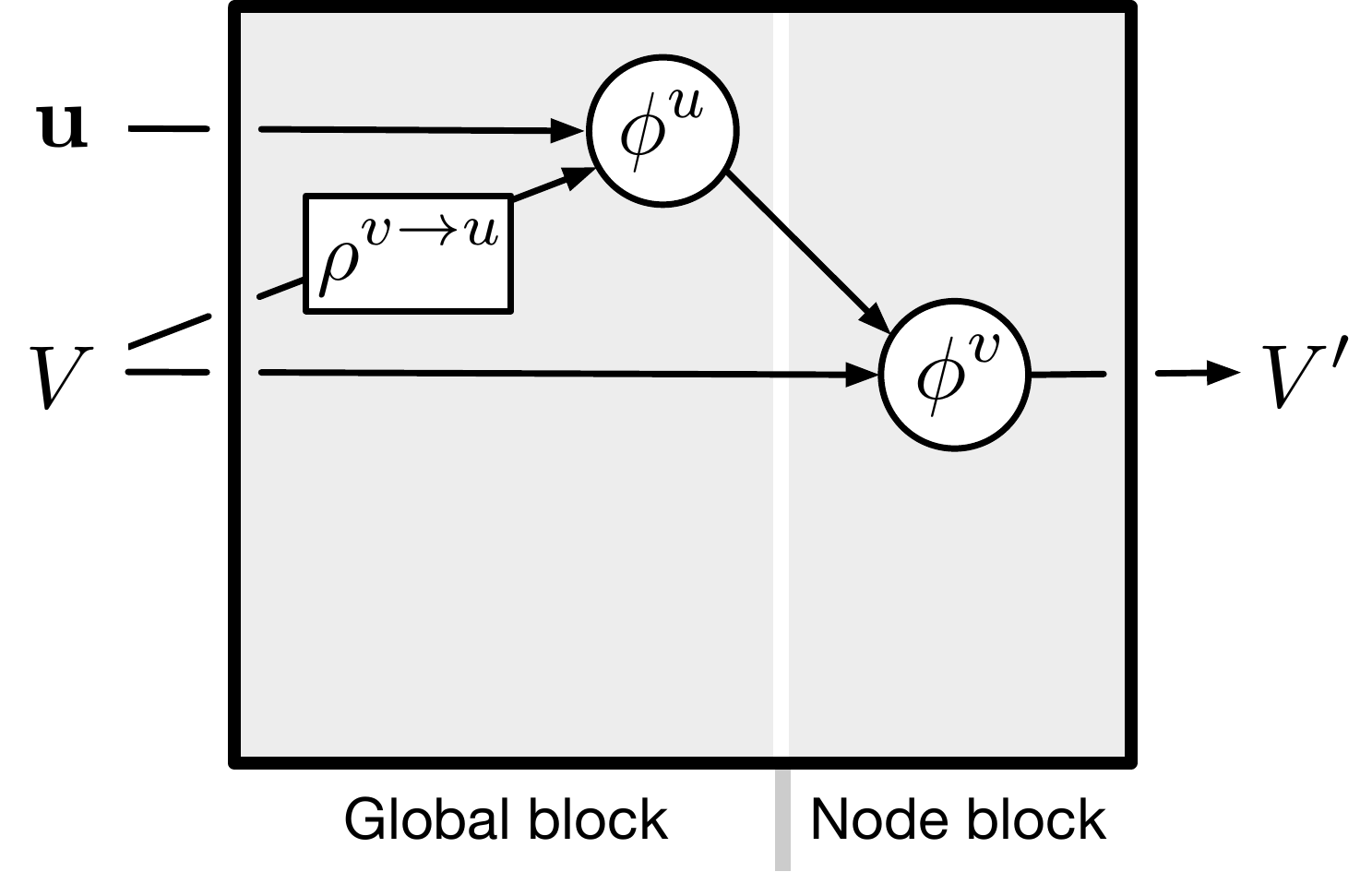}
		   }
	\caption{(a) The Graph Convolutional Network (GCN)~\cite{kipf2016semi}, a type of message-passing neural network, can be expressed as a GN, without a global attribute and a linear, non-pairwise edge function. (b) A more dramatic rearrangement of the GN's components gives rise to a model which pools vertex attributes and combines them with a global attribute, then updates the vertex attributes using the combined feature as context.}
	\label{fig:gcn}
\end{figure}

The GN formalism is a general framework which can capture a variety of other GNN architectures. Such architectures can be expressed by removing or rearranging internal components of the general GN block in Figure~\ref{fig:gn-framework}, and implementing the various $\phi$ and $\rho$ functions using specific functional forms. For example, one very popular GNN architecture is the Graph Convolutional Network (GCN)~\cite{kipf2016semi}. Using the GN formalism\cite{battaglia2018relational,gilmer2017neural}, a GCN can be expressed as,
\begin{alignat*}{2}
    \mathbf{e}'_k &= \phi^e\left(\mathbf{e}_k, \mathbf{v}_{s_k} \right) &&= \mathbf{e}_k \mathbf{v}_{s_k} \ , \qquad\text{where } \mathbf{e}_k = \frac{1}{\sqrt{\text{degree}(r_k)\text{degree}(s_k)}} \\
    \mathbf{\bar{e}}'_i &= \rho^{e\rightarrow v}\left(E'_i\right) &&= \sum_{\{k \, \vert \, r_k = i\}} \mathbf{e}'_k \\
    \mathbf{v}'_i &= \phi^v\left(\mathbf{\bar{e}}'_i \right) &&= \sigma \left(\mathbf{\bar{e}}'_i W\right)
\end{alignat*}
Figure~\ref{fig:gcn} shows the correspondence between the GCN and the GN depicted in Figure~\ref{fig:gn-framework}.

In section~\ref{sec:guidelines} we will discuss the considerations taken into account when deciding how to choose the actual implementation of the GNs internal functions. The choice of the specific architecture is motivated by the relationships that exist between the elements in the input data and the task one is trying to solve with the model.

\section{Survey of Applications to Particle Physics}\label{sec:applications}

Beyond discriminating signals from background in physics analysis, machine learning can be applied in many of the steps of the event: triggering, reconstruction and simulation.
GNNs are used in three different ways to make predictions : at the level of the graph, or node, or edge, depending on the task at hand.
We described briefly below the challenges and the methods applied, coming back in further details in~\ref{sec:guidelines}.
All the presented methods were developed on simulated events, and no performance on real data is reported so far.

In each line of work described below, a decision was first made about how the data could be expressed as a graph: What are the entities and relations which would be represented as nodes and edges, respectively? What is the required output, i.e., edge-, node-, or graph-level predictions? From there, choices about the specific GNN architecture were made to reflect the desired computation: Is a global output network required to produce graph-level outputs? Should pairwise interactions among nodes be computed, or more GCN-like summation and non-linear transformation? How many message-passing steps should be used, in order to propagate information among distant nodes in the graph?

\subsection{Graph Classification}

\paragraph{Jet Classification.}
\textit{Jets} or \textit{showers} are sprays of stable particles that are stemming from multiple successive interaction and decays of particles, originating from a single initial object.
The identification of this original object is of paramount importance in particle physics.
Because of the rather large lifetime of the b-hadrons~\cite{Tanabashi:2018oca} and hence a significantly displaced decay vertex, identification of b-jet (\textit{b-tagging}) using classical methods has been rather successful.
With the advent of deep learning methods, lower level information has been used to improve the performance of b-tagging, and opened the possibility of identifying jets coming from other particle (c-hadron, top-quark, tau, etc).
The jets coming from pure hadronic interaction driven by quantum chromo-dynamics (QCD) (so called \textit{QCD jets}), are covering an extremely large phase space and constitute an irreducible background to other classes of jets.
In particular, within the framework of the particle flow reconstruction \cite{Sirunyan:2017ulk}, the event is interpreted through a set of particle candidates.
As such, in references \cite{henrionneural,Komiske_2019,qu2019particlenet,Moreno_2020,moreno2019interaction,mikuni2020abcnet,Bernreuther:2020vhm} the collection of particle candidates is represented on a graph and various methods are applied.

The authors of~\cite{henrionneural} use a fully connected graph, and message passing architecture to learn the adjacency matrix, comparing several directed and undirected graph constructions.
The classification of jets originating from the hadronic decay of a $W$ boson and QCD jets is shown to improve with the proposed method.
Work on physics-based inductive biases is left for future work to improve the learning of the adjacency matrix.
It should be noted that learning the adjacency matrix is related to learning attention in~\cite{mikuni2020abcnet}.

In~\cite{qu2019particlenet} the authors use the \textit{edgeconv} method from~\cite{wang2018dynamic} to derive a point cloud architecture for jet tagging.
The connectivity of the graph is defined dynamically by computing node neighborhoods over the distance in either the input space, or an intermediate latent space when graph layers are stacked.
The architecture respects the particle permutation invariance by mean of averaging of contributions from the connected neighbors. 
The performance of this model for the quark/gluon discrimination (separating jets originating from a quark or a gluon) and top tagging (discriminating hadronic top decay and $QCD$ jet) tasks is reported to be better than other previously studied architectures. The learned edge function is constrained to taking as input a node feature and the feature difference between this node and the connected node.
In \cite{Bernreuther:2020vhm} the same model architecture is applied to the specific case of semi-visible jet originating from the cascade decay of hypothetical dark hadrons. 
The method outperforms neural networks that operate on images, as well as models including physical inductive biases \cite{Butter:2017cot}.
The authors demonstrate an order of magnitude improvement on the sensitivity of dark matter search when using this method.

The authors of~\cite{Moreno_2020,moreno2019interaction} take inspiration from~\cite{battaglia2016interaction} and adapt the interaction network architecture to the purpose of graph categorisation.
Using a fully connected graph over the particles of a jet and primary vertices of the event, a graph category is extracted after one step of message passing.
The performance of this model on a multi-class categorisation (light quarks, gluon, W and Z bosons hadronic decays, and hadronic top jets) is better than other non-graph-based architectures against which it was compared.
On the specific use case of tagging jets which stem from Higgs bosons decaying onto a pair of b quarks, the algorithm outperforms state of the art methods, even when the proper mass decorrelation method \cite{ATLAS:2018ibz} is applied.
The authors report some potential computation performance issues with running the model for predictions.
The measurement however, is done with a model obtained from a format conversion between major frameworks, and the performance could be improved with a native implementation instead.

With~\cite{Komiske_2019} the authors applied the \textit{Deep Sets} method from~\cite{NIPS2017_6931} to jet tagging.
They propose a simplified model architecture with provable physics properties, such as infrared and colinear safety.
The features of each particle are encoded into a latent space and the graph category is extracted from the summed representation in that latent space.
The model has no connectivity, and thus no attention or message passing, and pools information globally across all the elements before the categorisation is output, and yet the performance of this simple model on the quark/gluon classification is surprisingly on par with other more complicated models.
The authors provide ways of interpreting what the model has learned, and are able to extract closed-form observables from their trained model.

In~\cite{mikuni2020abcnet} the \textit{graph attention network} from~\cite{velickovic2018graph} is adapted for graph categorisation.
The node and edge features are created and updated by means of multiple fully connected neural networks, operating on the graph, and an additional attention factor, equivalent to a weighted, directed adjacency matrix is computed per directed edge, and used in the update rule.
A k-nearest neighborhood connectivity pattern is constructed using the distance over the edge features, initialized to the difference between node features, and later in a latent space when using stacked graph layers.
Stability of the models is improved with the use of a multi-head mechanism, and skip connections at multiple level are added to facilitate the information flow.
Their model outperforms the model from~\cite{qu2019particlenet} on the quark/gluon classification task, indicating the importance of the attention mechanism --- to which we come back to in sections~\ref{sec:modelarch} and~\ref{sec:discussion}.

\paragraph{Event Classification.}
Here we use the term \textit{event} for the capture by an experiment of the full history of a physics process.
In astroparticle, for example it is the collection of signals that covers the interaction of an high energy particle interacting with the atmosphere. 
The jet tagging task presented in the previous section is part of a full event identification in collider physics.
Event classification is the task of predicting or inferring the physics process at the origin of the recorded data.

The authors of~\cite{choma2018graph} applied a graph convolution method for the classification of the signal in the IceCube detector, to determine if a muon originated from a cosmic neutrino, or from a cosmic ray showering in the earth atmosphere.
The adjacency matrix of a fully connected graph of the detector sensors is constrained to a Gaussian kernel on the physical distance, with a learnable locality parameter.
Node features are updated by application of the adjacency matrix and non-linear activation.
The graph property is extracted from the sum over the latent features of the nodes of the graph.
This GNN model yields a signal-to-background ratio about three times as big as the baseline analysis of such signal.

In~\cite{Abdughani:2018wrw}, the \textit{message passing neural network} architecture from~\cite{gilmer2017neural} is used over a fully connected graph composed of the final state particles, and the missing transverse energy.
Messages are computed from the node features and a distance in the azimuth-rapidity plane first, then in the node latent space for later iterations.
Such messages are passed across the graph in two iterations, and each node receives a categorisation.
The node-averaged value is used to predict the event category.
The model is compared to densely connected models, and is showing superior performance when comparing the $S/\sqrt{B}$ analysis significance.
From the same authors, in~\cite{Ren:2019xhp,Abdughani:2020xfo}, a similar architecture is applied to event classification for other signal topologies, demonstrating the versatility of the method.

\subsection{Node Classification and Regression}

\paragraph{Pileup Mitigation.}

In a view to increase the overall probability of producing rare processes and exotic events, the particle density of bunches composing the colliding beams can be increased.
This results in multiple possible interactions per beam crossing. 
The downside of this increased probability is that, when occurring, an interesting interaction will be accompanied with other spurious, less interesting interactions (\textit{pileup}), considered as noise for the analysis.
Mitigation of pileup is of prime importance for analysis at colliders.
While it is rather easy to suppress charged particles by virtue of the primary vertex they are originating from, neutral particles are harder to suppress.
In a particle flow reconstruction \cite{Sirunyan:2017ulk}, the state of the art is to compute a pileup weight per particle \cite{Bertolini:2014bba}, and use it for mitigation.

In~\cite{martinez2018pileup} the authors utilize the \textit{gated graph network architecture}~\cite{li2015gated} to predict a per particle probability of belonging to the pileup part of the event.
The graph is composed of one node per charged and neutral particle in the event, and the connectivity is imposed to $\Delta R \equiv \sqrt{\delta \phi^2 + \delta \eta^2} < 0.3$ in the azimuth-pseudorapidity plane.
An averaged R-dependent message is computed and gated with each previous node representation by mean of a gated recurrent unit (GRU) to form the new node representation.
The per-particle pileup probability is extracted with a dense model, after three stacked graph layers, and a skip connection into the last graph layer.
The model outperforms other standard methods for pileup subtraction and improves resolution of several physical observables.

The authors of~\cite{mikuni2020abcnet} take inspiration from the \textit{graph attention network} from~\cite{velickovic2018graph} to predict a per-particle pileup probability.
An architecture very similar to the one used for the jet classification (described previously) is used to create a global graph latent representation, which in turn is used to compute an output that is mapped back to each node, thanks to a given order of the latter.
This method is shown to improve the resolution on the jet and di-jet mass observables, while being stable over a large range of pileup density.

\paragraph{Calorimeter Reconstruction.}

A \textit{calorimeter} is a detector which goal is to contain and measure the total energy of a system.
In particle physics, a calorimeter is commonly composed on the one hand of inactive material inducing showering of particles and energy loss (\textit{absorber}), and on the other hand a sensitive material that aims at measuring the collective released energy in the absorber.
Reconstruction of the energy of the incoming particle in such a sampling calorimeter involves calibration and clustering of the signal of various cells.

With~\cite{Qasim_2019} a graph network based approach is proposed to cluster and assign the signal in a high granularity calorimeter to two incoming particles.
A latent edge representation is constructed in the latent space of the nodes, using a potential function of the distance also in the latent space.
Two methods are proposed for the graph connectivity, one --- \textit{GravNet} --- using nearest neighbors in a latent space, the other --- \textit{GarNet} --- using a fixed number of additional nodes (dubbed \textit{aggregator}) in the graph.
Node features are updated using concatenated message from multiple aggregation methods, and provides in output the fraction of energy of the cell belonging to each particle.
The proposed methods are slightly improving over more classical approaches, and could be beneficial in more complex detector geometry than the one studied.

\paragraph{Particle Flow Reconstruction.}
Typically, detectors in particle physics are composed of multiple sub-detectors with various sensing technologies.
Each sub-detector is targeting the measurement of specific characteristic of the particle.
The assembly of all measurements allows for the characterisation of the particle properties.
The \textit{particle flow} --- or \textit{energy flow} --- reconstruction is an algorithm that aims at assigning to a candidate particle all the measurements in each sub-detector \cite{Sirunyan:2017ulk}.
Since all particles produced during a collision can potentially be reconstructed, \textit{particle flow reconstruction} allows for fine grained interpretation and analysis of collision events.

The author of~\cite{kieseler2020object} proposes the \textit{object condensation} loss formulation, using a GNN method to extract the particles' information from the graph of individual measurements.
In this context, the model is set to predict the properties of a smaller number of particles than there are measurements, in essence doing a graph reduction.
A stacked-\textit{GravNet}-based model performs node-wise regression of a kinematic corrective factor together with a \textit{condensation weight}.
The latter indicates whether a node of the graph has to be considered as representative of a particle in the event, and have its regressed quantities be assigned to that particle.
The performance of this algorithm is compared with a baseline particle-flow algorithm on rather sparse large hadron collider (LHC) environments.
The proposed method is shown to be more efficient and produces less fake particles than the standard approach.

\paragraph{Efficiency Parametrization.}
The analysis of particle physics data --- in particular collider experiment data --- requires applying selection criteria on the large volume of data, in a view to enhance the proportion of interesting signals.
It is crucial to determine with as little uncertainty as possible the fraction of signal passing these selections, if one wants to measure the rate of production of that signal during the experiment.
Much care is taken to determine these selection efficiencies, as they play significant roles in measuring the cross section of known processes, or while setting limits on production of unknown signals.
The efficiencies can be measured from data or simulation, per event or any component of it.
It is often the case that the efficiency of a specific selection on a component of the full events also depends on the other components of the event.
Taking into account the correlation between all components of an event is a hard task that machine learning can help with.

The authors of~\cite{badiali2020efficiency} use GNNs to learn the per-jet tagging efficiency, from a fully connected graph representation of the jets in the event.
The model is a message passing GNN. The edge update and node updates are both implemented as simple fully connected networks. The final node representation is used to predict the per-jet efficiency for each jet in an event.
The GN allows taking into account the dependency of the per-jet efficiency on the other jets in the event.
The comparison is made with the classical method of explicitly parametrizing the per-jet efficiency with a two dimensional histogram, whose axis are the jet transverse momentum and pseudo-rapidity.
The authors show how the graph representation and GNN parametrisation allows improving determination of the per-jet efficiency, compared to the more traditional method.

\subsection{Edge Classification}

\paragraph{Charged Particle Tracking.}
Charged particles have the property of ionizing the material they traverse.
This property is utilized in a tracking device (\textit{tracker}) to perform precise measurement of the passage of charged particles.
Contrary to calorimeters, trackers should not alter too much the energy of the incoming particle, as such it usually produces a sparse spatial sampling of the trajectory.
The reconstruction of the trajectory of original particles amounts to finding what set of isolated measurement (\textit{hits}) belong to the same particle.
Most tracking devices are embedded in a magnetic field that will curve the trajectories and hence provide a handle at measuring the particle momentum component transverse to the magnetic field, since this quantity and the curvature are inversely proportional.

The authors of~\cite{farrell2018novel} propose a GNN approach to charged particle tracking using edge classification.
Each node of the graph represents one sparse measurement, or hit, with edge constructed between pairs of hits with geometrically plausible relations.
Using multiple updates of the node representation and edge weight over the graph (using the edge weight as attention), the model learns what are the edges truly connecting hits belonging to the same track.
This approach transforms the clustering problem into an edge classification that defines the sub-graphs of hits belonging to the same trajectory.
The performance of this method has high accuracy when applied in a simplified case, and is promising for more realistic scenarios.
In~\cite{Ju:2020xty}, a GNN model involving message passing is presented and provides improved performance.

\paragraph{Secondary Vertex Reconstruction.}
The particles within a jet often originate from various intermediate particles that are worth identifying for the purpose of identifying the origin of the jet (see the paragraph on jet identification above).
The decay of the intermediate particles are identified as secondary vertices within the jet, using clustering algorithms on the particles, such as the adaptive vertex reconstruction~\cite{5734880}.
Based on the association to secondary vertex, the particles within a jet can henceforth be partitioned.

In~\cite{serviansky2020set2graph}, the authors develop a general formalism for \textit{set-to-graph} neural networks and provide mathematical proof that their formulation is a universal approximation of function mapping a graph structure onto an input set --- all invariance taken into account.
In particular, they apply a \textit{set-to-2-edge} --- predicting single edge characteristics from the input set --- approximation to the problem of particle association within a jet.
The model is a composition of an embedding model, a fixed broadcasting mapping and a graph-to-graph model.
All components are actually rather simple and the expressivity of the full model stems from the specific equivariant formulation.
Their model outperforms the standard methods on jet partitioning by about 10\% over multiple metrics.

\section{Formulating HEP tasks with GNN}\label{sec:guidelines}

The articles described in section~\ref{sec:applications} make use of multiple graph connectivity schemes, model architecture and loss functions. 
Experience shows that using our knowledge about the underlying physics in order to encode the relationship between the nodes --- whatever they may represent --- in both the input graph and the model architecture is key in developing algorithms. 
Unfortunately it is not always clear which methods and model architectures will outperform the others. 
This section aims to clarify the choices made and provide a checklist of considerations for the particle physicist looking to develop a new application using a GNN. 

\subsection{Task Definition}

The first step is to decide what function one wants to learn with the GNN. 
In some applications this is trivial - for example jet, event or particle classification. 
In those cases a GNN is used to learn some representation of the node or the entire graph/set and a standard classifier is trained on that representation.

For tasks such as segmentation or clustering, there is a choice between formulating the task as edge classification or something like the object condensation method which uses node representations to formulate a partition of the input set. The object condensation method has an important advantage, in that it computes relationships between objects (the attractive or repulsive potential) only while training the algorithm, in the computation of the loss function. An edge classifier will learn an edge representation and use that to classify edges. The number of edges can be large, increasing the computation and memory requirements of the algorithm. The determination of the set partition in the object condensation method is a simple function of the node representation, which greatly reduces those requirements.

None of the work presented in section~\ref{sec:applications} is using a mapping of the input onto the edges of the graph. 
Because an edge can only link two nodes --- while a node can be connected to as many edges as desirable --- construction of such graph would require a specific structure of the input.
One such use case could be in situations where observations arise from two concurrent measurements, such as hit position in stereo strip detectors.
The detector is composed of two rectangular modules with a thin strip of sensors along one dimension, and the modules are tilted with respect to each other by a couple of degrees so as to have the strip sensors overlapping and hence creating a grid.
With strip measurement positioned on the nodes, the important information would be located on the edges, as a combination of two such hits.
Other examples in network communication might also be relevant.

\subsection{Graph Construction}\label{sec:graphconstruction}

\begin{figure}[b]
	\centering
	\subfigure[]{
	\includegraphics[]{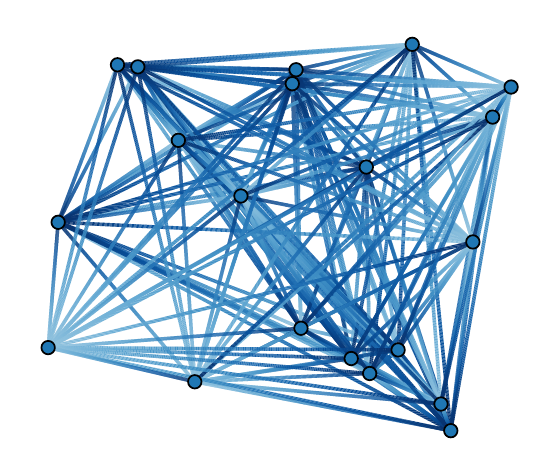}
	}
	\subfigure[]{
	\includegraphics[]{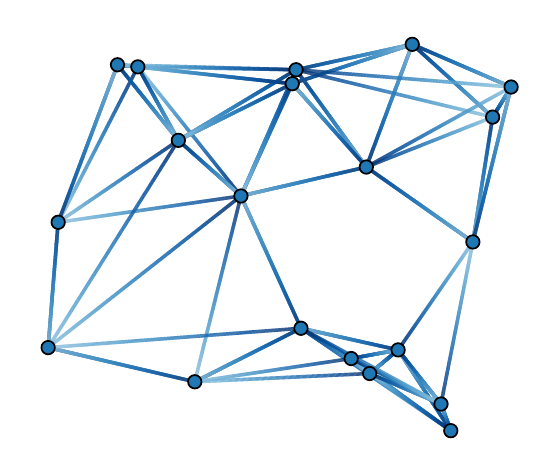}
	}
	\subfigure[]{
	\includegraphics[]{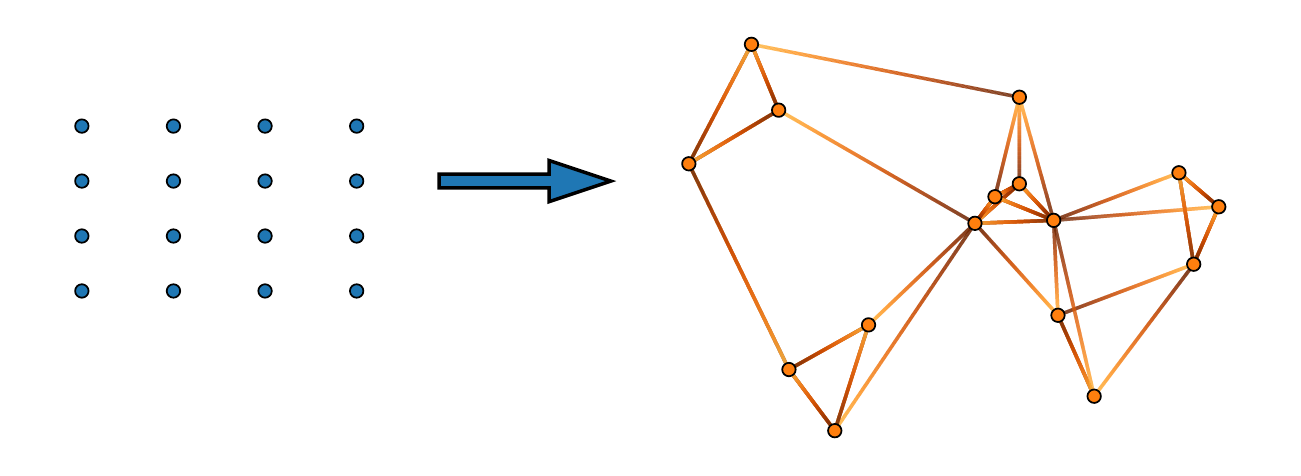}
	}
	\caption{Different methods for constructing the graph. (a) Connecting every node to every other node (b) Connecting neighboring nodes in some predefined feature space (c) Connecting neighboring nodes in a learned feature space.}
	\label{fig:graphconstruction}
\end{figure}

In most particle physics applications, the nature of the relationships between different elements in the set are not clear cut (as it would be for a molecule or a social network). Therefore a decision needs to be made about how to construct a graph from the set of inputs. Different graph construction methods are illustrated in figure~\ref{fig:graphconstruction}.
Depending on the task, one might even want to avoid creating any pairwise relationships between nodes. If the objects have no pairwise conditional dependence --- a DeepSet~\cite{Komiske_2019}  architecture with only node and global properties might be more suitable.
Edges in the graph serve 3 roles: 
\begin{enumerate}
	\item The edges are communication channels among the nodes.
	\item  Input edge features can indicate a relationship between objects, and can encode physics motivated variables about that relationship (such as $\Delta R$ between objects).
	\item Latent edges store relational information computed during message-passing, allowing the network to encode such variables it sees relevant for the task.
\end{enumerate} 

In cases where the input sets are small ($N_v \sim \mathcal{O}(10)$ ) the typical and easiest choice is to form a fully connected graph, allowing the network to learn which object relationships are important.
In larger sets, as the number of edges between all nodes increases as $N_e \propto ({N_v})^2$, the computational load of using a neural network to create an edge representation or compute attention weights becomes prohibitive. One possible work-around is to choose a fixed edge feature that is easy to pre-compute --- such as distance between detector modules.

If an edge-level computation is required, it is necessary to only form some edges. Edges can be formed based on a relevant metric such as the $\Delta R$ between particles in a detector, or the physical distance between detector modules. Given a distance measure between nodes, some criterion for connecting them needs to be formulated, such as connecting k-nearest neighbors in the feature space. 

The node features used to connect edges can also be based on a learned representation. This is sometimes referred to as \textit{dynamic} graph construction, and used by the EdgeConv~\cite{qu2019particlenet} and GravNet~\cite{Qasim_2019} architectures, for example.
We will discuss this in more detail in section~\ref{sec:modelarch}, showing the connection between the idea of dynamic graph construction and attention mechanisms.

When the graph is constructed dynamically, such as using the node representation to connect edges between k-nearest neighbors, the gradient of the neural network parameters is only affected by those nodes that have actually been connected.
Since the indexing of node-neighborhood is non differentiable, its parameters cannot be learn with gradient descent, but can be optimized on hyper-parameter search.

In initial stages of the training, the edge formation is essentially random, allowing the network to explore which node representations should be closer together in the latest space. 
During later stages of the training, one may wish to encourage further exploration by the network. 
One possible way to do this is to inject random edges --- for example besides connecting nodes to k-nearest neighbors in latest space, connecting an additional small number of random connections to nodes further away in the latent space.

A recent paper~\cite{johnson2020learning} introduces a reinforcement learning agent which traverses an input graph to reach nodes which should be connected by new edges. Its policy is optimized for some downstream task performance, so that the nodes it chooses to connect with new edges improve the task performance.

\subsection{Model Architecture}
\label{sec:modelarch}

Designing the model architecture should reflect a logical combination of the inputs towards the learning task. In the language of the GN formalism (section~\ref{sec:GNformalism}), we need to select a concrete implementation of the GN block update and aggregation functions $\phi$ and $\rho$, and decide how to configure their sequence inside the GN block. Additionally we need to decide which kinds of GN blocks we want to combine and how to stack them together.
As explained in section~\ref{sec:GNformalism}, different architectures such as Graph Convolution Networks, Graph Attention Networks, are specific choices for constructing a GNN --- but they are all equivalent in the sense that their output is a graph with learned node/edge/graph representations which are then used to perform the actual task.

\paragraph{GN block functions.}
The key question here is what logical steps one would take to form the GN block output in a way that serves the task, and which parts of this logical process should be modeled with neural networks?
The most general GN block (as shown in figure~\ref{fig:full_gn_block}) could have all of its update functions implemented as neural networks, which allows the most flexibility in the learning processes. This flexibility might not be required for the task, and it might carry computational costs that we wish to keep to a minimum. Therefore its probably better to start with a simple architecture, and only add complexity gradually, until the algorithms performance is satisfactory.

\begin{figure}
	\centering
	\includegraphics[]{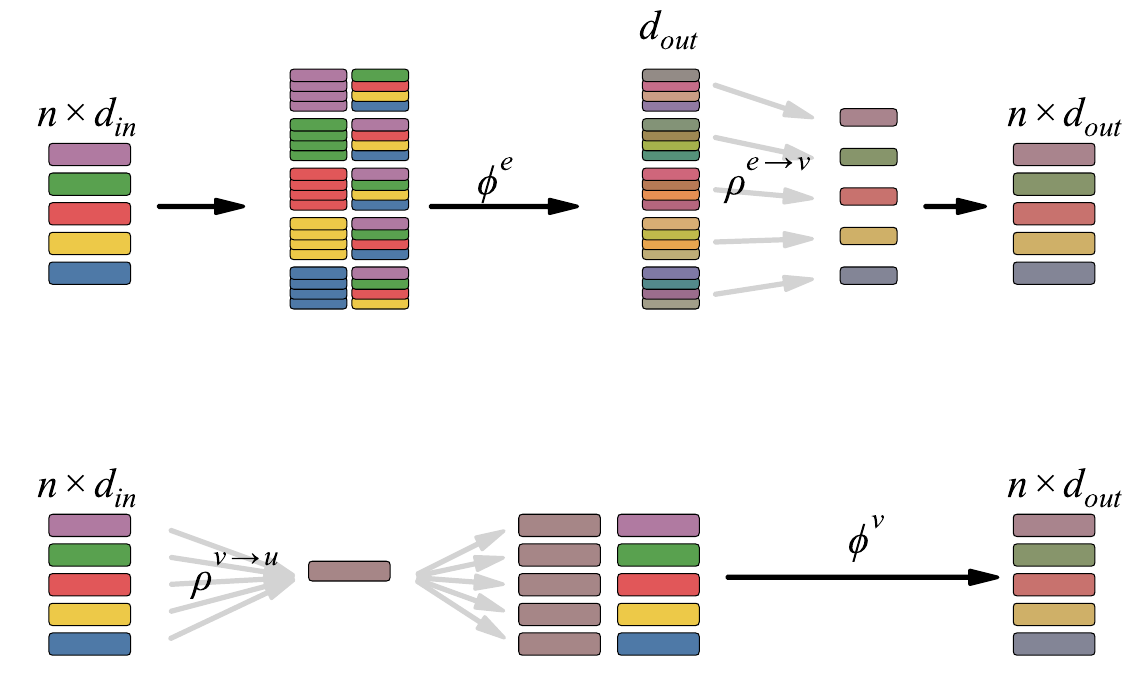}
	\caption{Possible architectures for a GN block that create an updated node representation. 
	Using an edge representation as an intermediate step (upper diagram) gives a different inductive bias to the model, compared to using a global representation of the set (lower diagram).
	The function names are from equation \ref{eq:gn-functions} and figure \ref{fig:full_gn_block}}
	\label{fig:choosingGNblockfunctions}
\end{figure}

Figure~\ref{fig:choosingGNblockfunctions} shows two examples of possible configurations, either creating an edge representation before aggregating edges and forming a node update, or using global aggregation before a node update. Both configurations result in an updated node representation, but one of them is based on a sum of pair-wise representations, and the other on a global sum of node representations --- the information content is the same, but the inductive bias is different. 
For example, the authors of~\cite{badiali2020efficiency} assumed that the jet-tagging efficiency is heavily affected by the $\Delta R$ between neighboring jets --- therefore an edge update step created a representation of pair-wise interaction between jets, which was then summed for each jet to create the updated node representation. 
In contrast the authors of~\cite{Komiske_2019} used a DeepSet architecture, where each node representation is created independently from its neighbors, the node representations are then summed to create the graph representation, with each node representation weighted by the particles energy.

\paragraph{Attention Mechanisms.}

\begin{figure}
	\centering
	\includegraphics[]{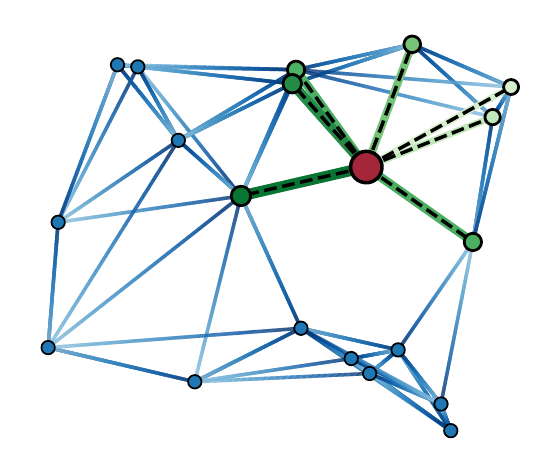}
	\caption{Attention mechanisms allow the network to learn relative importance of different nodes/edges in the aggregation functions. The red node is a node whose neighbors are being aggregated by $\rho^{e\rightarrow v}$ , the attention mechanism will learn to provide relative weights for the adjacent nodes/edges (the green highlights) such that the output of $\rho^{e\rightarrow v}$ is a weighted sum of either the node or edge representations.}
	\label{fig:attention}
\end{figure}

Another important component that can be used in defining the $\rho^{e\rightarrow v}$ and $\rho^{v,e\rightarrow u}$ aggregation functions is using \textit{attention mechanisms}, as illustrated in figure~\ref{fig:attention}. The term \textit{attention} is rooted in the perceptual psychology and neuroscience literatures, where it refers to the phenomenon and mechanisms by which a subset of incoming sensory information is selected for more extensive processing, while other information is deprioritized or filtered out.

The key consideration for defining and adding an attention mechanism is whether different parts of the input data are more important than others.
For example, in classifying jets, some particles that originate from a secondary decay are an important footprint of a particular class of jets --- therefore those particles may be more important for the classification task.
There are a few different implementations of attention mechanisms. They all share the basic concept of using a neural network or a pre-defined function to compute weights which represent the relative importance of different elements in a set. In the GN block $\rho$ functions, these weights are used to create weighted sums of the representations of the different elements. 

\begin{figure}
	\centering
	\includegraphics[]{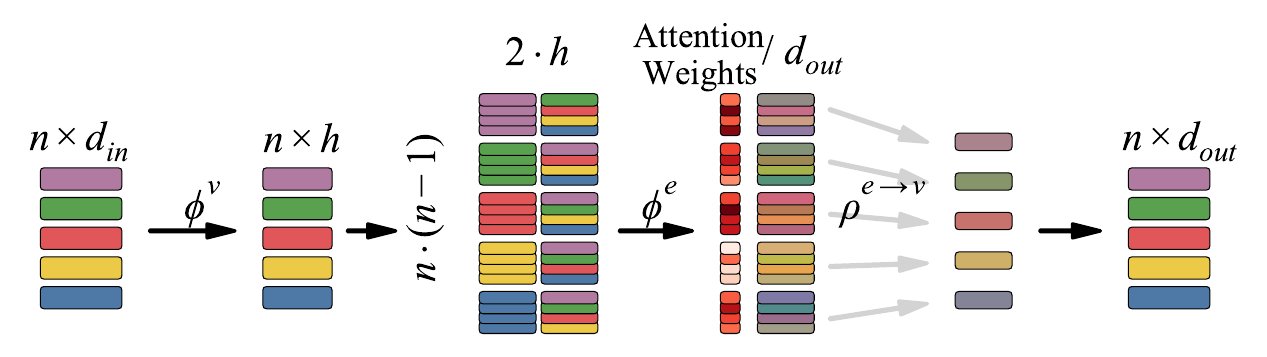}
	\caption{The GN block structure of the EdgeConv, GravNet layers as described in the GN formalism. The node embedding stage is a GN block which operates on the nodes independently (without any information exchange between them), followed by a GN block which creates an edge representation for every pair of vertices, aggregates edges for each node and then updates the vertex representation. The edge update function $\phi^e$ does not use a neural network, but uses a pre-defined function of the node representation - leading to a reduction in computational cost.}
	\label{fig:compareEdgeConvGarNet}
\end{figure}

Here we want to draw attention to the connection between attention mechanisms and dynamic graph construction. Figure~\ref{fig:compareEdgeConvGarNet} shows the structure of two architectures discussed in section~\ref{sec:applications}, the EdgeConv, GravNet layers. These are both GN block implementations, they take as input a set of nodes (without explicit edges) and output an updated node representation. Both begin with a node embedding stage, which creates a node representation without exchanging information between the nodes. This node embedding (or only part of its feature vector, in the case of GravNet) is interpreted as a position of the node in a latent euclidean space, and edges are formed between k-nearest neighbors. This can be thought of as a fully connected graph with an attention mechanism that assigns a weight of 1 to nodes within the set of k-nearest neighbors, and 0 otherwise. The advantage of this procedure over using a neural network to compute attention weights is the much lower computational cost of both computing the edge attention weight and the subsequent edge-related operations. 

Its worth noting that the GarNet layer~\cite{Qasim_2019} can be described as a form of \textit{multi-headed} self-attention mechanism~\cite{vaswani2017attention}. The GarNet layer interprets the node embedding as $s$ different ``distances'' (with $s$ being the dimension of the embedding). These distances are attention weights over each node of the graph, and they are used to compute $s$ different weighted sums --- these are the $s$ different \textit{heads} of the attention mechanism. 
The weighted sums are propagated back to the nodes again via attention weights of each node to each of the $s$ attention heads. The reason GarNet is computationally affordable without a hard cutoff --- such as k-nearest neighbors --- is that $\phi^v$, the node embedding function, is the only one computed with a neural network. The attention weights are all computed with pre-defined functions given the node embedding (specifically, the function is $exp(-|w|)$ where $w$ is the attention weight).

\paragraph{Stacking GN blocks.}

A stack of GN blocks (as described in figure~\ref{fig:gn_stack}) serves two purposes. 
First, in the same way that stacked layers in any neural network architecture (such as a CNN) can be thought of as gradually constructing a high level representation of the data, GN blocks arranged sequentially serve the same purpose for constructing the node/edge and graph representations. Therefore, additional GN blocks increase the depth of the model and its expressive power.

\begin{figure}
	\centering
	\includegraphics[width=0.5\textwidth]{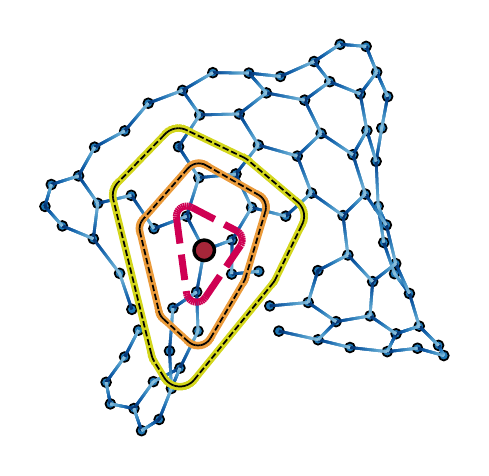}
	\caption{Each iteration of message passing between nodes increases a nodes receptive field. For example the node in red communicates with its three connected neighbors (red outline) in the first message passing step. The orange and yellow dotted lines represent the nodes that communicate after two and three iterations respectively. The node left out of the yellow line have not exchanged information with the red node, after three iterations only.}
	\label{fig:message_passing}
\end{figure}

Second, after one iteration of message passing in a single GN block, the node has only exchanged information with its immediate connected neighbors. This is illustrated in figure~\ref{fig:message_passing}. Multiple iterations with a GN block (either the same block applied multiple times, or different blocks applied in a sequence) increase each nodes receptive field, as the representation of its neighboring nodes was previously updated with information from their neighbors. Often skip or residual connections, which combine the input with the output, are used to prevent corruption of the updated representations, and preservation of the gradient signal, over many message passing steps, as is common in CNNs and RNNs.

\section{Summary and Discussion}\label{sec:discussion}

The papers reviewed in section~\ref{sec:applications} can be seen as the first wave of application of graph neural network architectures to diverse tasks in high energy physics.
The methods show superior performance over other model architecture, thanks to the inductive bias, reduction of number of parameters, more elaborated loss function, and above all a much more natural data representation.
Graphs are constructed from observable in various ways, often with sparse connectivity to lessen computational requirements.

While multiple architectures are presented with different names, and slightly different formalisms, they all share the core concept of exchanging information across the graph.
We deciphered the variety of models in section~\ref{sec:guidelines} by providing some considerations on how the models were build.
We provide in the following some new directions to be considered as future direction for the next generation of graph neural network applications in high energy physics.

\paragraph{Transformer, Reformer, etc.}
Following the discussion of the GravNet and EdgeConv layers in section~\ref{sec:modelarch} and their relation to attention mechanisms, another class of models which are closely related to GNNs, and which perform a type of soft structural prediction, are Transformer architectures, based on the self-attention mechanism~\cite{vaswani2017attention}. 
In GNN language, a Transformer computes normalized edge weights in a complete graph (i.e., a  graph with edges connecting all pairs of nodes), and passes messages along the edges in proportion to these weights, analogous to a hybrid of graph attention networks~\cite{velickovic2018graph} and GCNs~\cite{kipf2016semi}. 

In GN notation, described in~\cite{battaglia2018relational} and used explicitly in graph attention networks~\cite{velickovic2018graph}, the Transformer uses a $\phi^e$ which produces both a vector message and a scalar unnormalized weight, and the $\rho^{e\rightarrow v}$ function normalizes the weights before computing a weighted sum of the message vectors. This allows a set of input items to be treated as nodes in a graph, without observed input edges, and the edge structure to be inferred and used within the architecture for message-passing. Different variants of attention mechanisms are a way to give different weights in the pooling operations $\rho^{e\rightarrow v}$, $\rho^{v,e\rightarrow u}$, as illustrated if figure~\ref{fig:attention}. The implementation of attention should reflect the nature of the interaction between the objects in the set, as they relate to the task.

The Reformer~\cite{kitaev2020reformer} architecture overcomes the quadratic computational and memory costs that challenge traditional Transformer-based methods, by projecting nodes into a learned high-dimensional embedding space where nearest neighbors are efficiently computed to inform a sparse graph over which to pass messages. The recent Linformer~\cite{wang2020linformer} method is similar, but with a low rank approximation to the soft adjacency matrix.

\paragraph{Graph generative models.}

\begin{figure}
	\centering
	\includegraphics[]{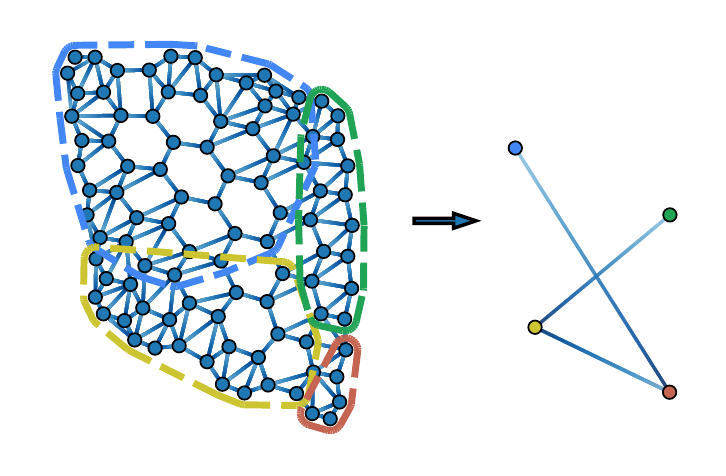}
	\caption{The DiffPool~\cite{ying2018hierarchical} layer and similar architectures allow to modify the graph structure as an intermediate step of the model computation. In the illustration, nodes in the input graph are grouped together to form nodes in the output graph. Each node is colored according to the outline of the nodes associated to it in the input graph. The output graph adjacency matrix is also learned as part of the DiffPool layer output. }
	\label{fig:graph_reduction}
\end{figure}

Importantly, the GN does not predict structural changes directly. However, many recent papers use GNs (or other GNNs) to decide how to modify a graph's structure. 
For example, \cite{li2018learning} and \cite{nash2020polygen} are autogressive graph generators, which use a GN or Transformer to predict whether a new vertex should be added to a graph (by the graph-level output), and which existing vertices to connect it to with edges (by the vertex-level outputs). 
The GraphRNN~\cite{you2018graphrnn}, and Graphite~\cite{grover2018graphite} are generative models over edges that use an RNN for sequential prediction, and GraphGAN~\cite{wang2018graphgan} is an analogous method based on generative adversarial networks. 
\cite{kipf2018neural}'s Neural Relational Inference treats the existence of edges as latent random variables, and trains a posterior edge inference front-end via variational autoencoding. 
In \cite{hamrick2018relational} and \cite{bapst2019structured}, a GN is used to guide the policy of a reinforcement learning agent and build graphs that represent physical scenes. 
The DiffPool~\cite{ying2018hierarchical} architecture (illustrated in figure~\ref{fig:graph_reduction} is an attention-based soft edge prediction mechanism, but over hierarchies of graphs, where lower-level ones are pooled to higher-level ones.


Generative models of graphs have not been explored much in particle physics, though some unpublished work is on-going.
The need for computational resource for simulation in particle physics is almost as large as the requirements for event reconstruction. 
There is a breadth of efforts on using machine learning as surrogate simulators in particle physics.
For the reasons exposed in section~\ref{sec:intro} that data in particle physics can often be represented as graphs, it is natural to investigate the use of generative models using graphs as a possible solution.
Models under development are for example predicting energy deposition in the cells of a calorimeter or the particle candidates obtained from a particle flow reconstruction algorithm.
In all cases, the generated quantities are naturally represented as a set or graph, with fixed or variable size.

\paragraph{Computation Performance.}

An important consideration for building and efficiently training GNNs on hardware is whether to use dense or sparse implementations of the graph's edges. 
The number of edges in a graph usually defines the memory and speed bottleneck, because there are typically more edges than nodes and the $\phi^e$ function is applied the most times. 
A dense adjacency matrix supports fast, parallel matrix multiplication to compute $E'$, which, for example, is exploited in speed-efficient GCN- and Transformer-style models. 
The downside is that the adjacency matrix's memory footprint is quadratic in the number of nodes.
Alternatively, using sparse adjacency matrices allows the memory to scale linearly in the number of edges, which allows much larger graphs to be processed. 
But the sparse indexing operations required to implement sparse matrix multiplication can incur greater time costs than their dense counterparts --- this is an active area of development for both software and hardware acceleration.
However, sparse operations are a key bottleneck in current deep learning hardware, and should next generation hardware substantially improve their speed, this would potentially improve the relative advantage of sparse edge implementations of GNNs.
In a computing environment in HEP, one cannot expect to have access to dedicated accelerators (GPU, TPU, FPGA, etc) --- although work is going in the direction of building the infrastructure --- and one needs to keep into consideration the time for running the model in production.

\paragraph{Final Remarks.}

Neural networks that operate on sets are increasing in popularity in high energy physics tasks, both in event reconstruction, and in physics analysis. 
These neural networks are performing well in proof-of-concept studies, either surpassing or matching existing state of the art techniques. 
They have not yet been tested in the field with real detector data.
It is important to understand that all of the models use the same basic building blocks to perform their tasks, and the most important consideration in designing the architecture for these neural networks is to correctly model the nature of interaction between the objects in the input set. 
It is probably the best practice to start with a simple graph model and architecture then build up on additional complexity geared towards incorporating scientific understanding of the physical process at stake.

\section*{Acknowledgement}

We thank Thomas Keck for valuable feedback on the manuscript. J.S is supported by the NSF-BSF Grant 2017600 and the ISF Grant 125756 and partially supported by the Israeli Council for Higher Education (CHE) via the Weizmann Data Science Research Center.
J-R.V. is partially supported by the European Research Council (ERC) under the European Union's Horizon 2020 research and innovation program (grant agreement n$^o$ 772369) and  by the U.S. Department of Energy, Office of Science, Office of High Energy Physics under award numbers DE-SC0011925, DE-SC0019227 and DE-AC02-07CH11359.

Data sharing is not applicable to this article as no new data were created or analysed in this study.

\vspace{1cm}

\bibliography{GNNinHEP}

\begin{thebibliography}{10}

\bibitem{Radovic:2018dip}
Alexander Radovic, Mike Williams, David Rousseau, Michael Kagan, Daniele
  Bonacorsi, Alexander Himmel, Adam Aurisano, Kazuhiro Terao, and Taritree
  Wongjirad.
\newblock {Machine learning at the energy and intensity frontiers of particle
  physics}.
\newblock {\em Nature}, 560(7716):41--48, 2018.

\bibitem{Carleo:2019ptp}
Giuseppe Carleo, Ignacio Cirac, Kyle Cranmer, Laurent Daudet, Maria Schuld,
  Naftali Tishby, Leslie Vogt-Maranto, and Lenka Zdeborová.
\newblock {Machine learning and the physical sciences}.
\newblock {\em Rev. Mod. Phys.}, 91(4):045002, 2019.

\bibitem{Guest:2018yhq}
Dan Guest, Kyle Cranmer, and Daniel Whiteson.
\newblock {Deep Learning and its Application to LHC Physics}.
\newblock {\em Ann. Rev. Nucl. Part. Sci.}, 68:161--181, 2018.

\bibitem{Bourilkov:2019yoi}
Dimitri Bourilkov.
\newblock {Machine and Deep Learning Applications in Particle Physics}.
\newblock {\em Int. J. Mod. Phys. A}, 34(35):1930019, 2020.

\bibitem{Larkoski:2017jix}
Andrew~J. Larkoski, Ian Moult, and Benjamin Nachman.
\newblock {Jet Substructure at the Large Hadron Collider: A Review of Recent
  Advances in Theory and Machine Learning}.
\newblock {\em Phys. Rept.}, 841:1--63, 2020.

\bibitem{livingreview}
HEP Community.
\newblock A living review of machine learning for particle physics
  \url{https://iml-wg.github.io/HEPML-LivingReview/}, 2020.

\bibitem{cranmer2019learning}
Miles~D Cranmer, Rui Xu, Peter Battaglia, and Shirley Ho.
\newblock Learning symbolic physics with graph networks.
\newblock {\em arXiv preprint arXiv:1909.05862}, 2019.

\bibitem{cranmer2020discovering}
Miles Cranmer, Alvaro Sanchez-Gonzalez, Peter Battaglia, Rui Xu, Kyle Cranmer,
  David Spergel, and Shirley Ho.
\newblock Discovering symbolic models from deep learning with inductive biases.
\newblock {\em arXiv preprint arXiv:2006.11287}, 2020.

\bibitem{mou2014convolutional}
Lili Mou, Ge~Li, Lu~Zhang, Tao Wang, and Zhi Jin.
\newblock Convolutional neural networks over tree structures for programming
  language processing, 2014.

\bibitem{shen2018ordered}
Yikang Shen, Shawn Tan, Alessandro Sordoni, and Aaron Courville.
\newblock Ordered neurons: Integrating tree structures into recurrent neural
  networks, 2018.

\bibitem{bronstein2017geometric}
Michael~M Bronstein, Joan Bruna, Yann LeCun, Arthur Szlam, and Pierre
  Vandergheynst.
\newblock Geometric deep learning: going beyond euclidean data.
\newblock {\em IEEE Signal Processing Magazine}, 34(4):18--42, 2017.

\bibitem{gilmer2017neural}
Justin Gilmer, Samuel~S. Schoenholz, Patrick~F. Riley, Oriol Vinyals, and
  George~E. Dahl.
\newblock Neural message passing for quantum chemistry, 2017.

\bibitem{battaglia2018relational}
P.~Battaglia, Jessica~B. Hamrick, Victor Bapst, Alvaro Sanchez-Gonzalez,
  Vinicius Zambaldi, Mateusz Malinowski, Andrea Tacchetti, David Raposo, Adam
  Santoro, Ryan Faulkner, Caglar Gulcehre, Francis Song, Andrew Ballard, Justin
  Gilmer, George Dahl, Ashish Vaswani, Kelsey Allen, Charles Nash, Victoria
  Langston, Chris Dyer, Nicolas Heess, Daan Wierstra, Pushmeet Kohli, Matt
  Botvinick, Oriol Vinyals, Yujia Li, and Razvan Pascanu.
\newblock Relational inductive biases, deep learning, and graph networks, 2018.

\bibitem{zhou2018graph}
Jie Zhou, Ganqu Cui, Zhengyan Zhang, Cheng Yang, Zhiyuan Liu, Lifeng Wang,
  Changcheng Li, and Maosong Sun.
\newblock Graph neural networks: A review of methods and applications.
\newblock {\em arXiv preprint arXiv:1812.08434}, 2018.

\bibitem{wu2019comprehensive}
Zonghan Wu, Shirui Pan, Fengwen Chen, Guodong Long, Chengqi Zhang, and
  Philip~S. Yu.
\newblock A comprehensive survey on graph neural networks, 2019.

\bibitem{ATLAS:2017dfg}
{Quark versus Gluon Jet Tagging Using Jet Images with the ATLAS Detector}.
\newblock 7 2017.

\bibitem{Kasieczka:2017nvn}
Gregor Kasieczka, Tilman Plehn, Michael Russell, and Torben Schell.
\newblock {Deep-learning Top Taggers or The End of QCD?}
\newblock {\em JHEP}, 05:006, 2017.

\bibitem{Macaluso:2018tck}
Sebastian Macaluso and David Shih.
\newblock {Pulling Out All the Tops with Computer Vision and Deep Learning}.
\newblock {\em JHEP}, 10:121, 2018.

\bibitem{Andrews:2018nwy}
M.~Andrews, M.~Paulini, S.~Gleyzer, and B.~Poczos.
\newblock {End-to-End Physics Event Classification with CMS Open Data: Applying
  Image-Based Deep Learning to Detector Data for the Direct Classification of
  Collision Events at the LHC}.
\newblock {\em Comput. Softw. Big Sci.}, 4(1):6, 2020.

\bibitem{Lin:2018cin}
Joshua Lin, Marat Freytsis, Ian Moult, and Benjamin Nachman.
\newblock {Boosting $H\to b\bar b$ with Machine Learning}.
\newblock {\em JHEP}, 10:101, 2018.

\bibitem{ATLAS:2019fxb}
{Convolutional Neural Networks with Event Images for Pileup Mitigation with the
  ATLAS Detector}.
\newblock 2019.

\bibitem{hochreiter1997long}
Sepp Hochreiter and J{\"u}rgen Schmidhuber.
\newblock Long short-term memory.
\newblock {\em Neural computation}, 9(8):1735--1780, 1997.

\bibitem{cho2014learning}
Kyunghyun Cho, Bart Van~Merri{\"e}nboer, Caglar Gulcehre, Dzmitry Bahdanau,
  Fethi Bougares, Holger Schwenk, and Yoshua Bengio.
\newblock Learning phrase representations using rnn encoder-decoder for
  statistical machine translation.
\newblock {\em arXiv preprint arXiv:1406.1078}, 2014.

\bibitem{ATLAS:2017gpy}
ATLAS Collaboration.
\newblock {Identification of Jets Containing $b$-Hadrons with Recurrent Neural
  Networks at the ATLAS Experiment}.
\newblock 2017.

\bibitem{Sirunyan:2020lcu}
CMS Collaboration.
\newblock {Identification of heavy, energetic, hadronically decaying particles
  using machine-learning techniques}.
\newblock {\em JINST}, 15(06):P06005, 2020.

\bibitem{Louppe:2017ipp}
Gilles Louppe, Kyunghyun Cho, Cyril Becot, and Kyle Cranmer.
\newblock {QCD-Aware Recursive Neural Networks for Jet Physics}.
\newblock {\em JHEP}, 01:057, 2019.

\bibitem{DIPs}
{Deep Sets based Neural Networks for Impact Parameter Flavour Tagging in
  ATLAS}.
\newblock Technical Report ATL-PHYS-PUB-2020-014, CERN, Geneva, May 2020.

\bibitem{schmidhuber2015deep}
J{\"u}rgen Schmidhuber.
\newblock Deep learning in neural networks: An overview.
\newblock {\em Neural networks}, 61:85--117, 2015.

\bibitem{lecun2015deep}
Yann LeCun, Yoshua Bengio, and Geoffrey Hinton.
\newblock Deep learning.
\newblock {\em nature}, 521(7553):436--444, 2015.

\bibitem{tensorflow2015-whitepaper}
Mart\'{\i}n Abadi, Ashish Agarwal, Paul Barham, Eugene Brevdo, Zhifeng Chen,
  Craig Citro, Greg~S. Corrado, Andy Davis, Jeffrey Dean, Matthieu Devin,
  Sanjay Ghemawat, Ian Goodfellow, Andrew Harp, Geoffrey Irving, Michael Isard,
  Yangqing Jia, Rafal Jozefowicz, Lukasz Kaiser, Manjunath Kudlur, Josh
  Levenberg, Dan Man\'{e}, Rajat Monga, Sherry Moore, Derek Murray, Chris Olah,
  Mike Schuster, Jonathon Shlens, Benoit Steiner, Ilya Sutskever, Kunal Talwar,
  Paul Tucker, Vincent Vanhoucke, Vijay Vasudevan, Fernanda Vi\'{e}gas, Oriol
  Vinyals, Pete Warden, Martin Wattenberg, Martin Wicke, Yuan Yu, and Xiaoqiang
  Zheng.
\newblock {TensorFlow}: Large-scale machine learning on heterogeneous systems,
  2015.
\newblock Software available from tensorflow.org.

\bibitem{paszke2017automatic}
Adam Paszke, Sam Gross, Soumith Chintala, Gregory Chanan, Edward Yang, Zachary
  DeVito, Zeming Lin, Alban Desmaison, Luca Antiga, and Adam Lerer.
\newblock Automatic differentiation in pytorch.
\newblock 2017.

\bibitem{scarselli2008graph}
Franco Scarselli, Marco Gori, Ah~Chung Tsoi, Markus Hagenbuchner, and Gabriele
  Monfardini.
\newblock The graph neural network model.
\newblock {\em IEEE Transactions on Neural Networks}, 20(1):61--80, 2008.

\bibitem{gori2005new}
Marco Gori, Gabriele Monfardini, and Franco Scarselli.
\newblock A new model for learning in graph domains.
\newblock In {\em Proceedings. 2005 IEEE International Joint Conference on
  Neural Networks, 2005.}, volume~2, pages 729--734. IEEE, 2005.

\bibitem{almeida1987learning}
LB~ALMEIDA.
\newblock A learning rule for asynchronous perceptrons with feedback in a
  combinatorial environment.
\newblock In {\em Proceedings of the IEEE 1st international conference on
  neural networks, 1987}, pages 609--618, 1987.

\bibitem{pineda1987generalization}
Fernando~J Pineda.
\newblock Generalization of back-propagation to recurrent neural networks.
\newblock {\em Physical review letters}, 59(19):2229, 1987.

\bibitem{li2015gated}
Yujia Li, Daniel Tarlow, Marc Brockschmidt, and Richard Zemel.
\newblock Gated graph sequence neural networks, 2015.

\bibitem{kearnes2016molecular}
Steven Kearnes, Kevin McCloskey, Marc Berndl, Vijay Pande, and Patrick Riley.
\newblock Molecular graph convolutions: moving beyond fingerprints.
\newblock {\em Journal of computer-aided molecular design}, 30(8):595--608,
  2016.

\bibitem{battaglia2016interaction}
Peter~W. Battaglia, Razvan Pascanu, Matthew Lai, Danilo Rezende, and Koray
  Kavukcuoglu.
\newblock Interaction networks for learning about objects, relations and
  physics, 2016.

\bibitem{sanchez2018graph}
Alvaro Sanchez-Gonzalez, Nicolas Heess, Jost~Tobias Springenberg, Josh Merel,
  Martin Riedmiller, Raia Hadsell, and Peter Battaglia.
\newblock Graph networks as learnable physics engines for inference and
  control.
\newblock {\em arXiv preprint arXiv:1806.01242}, 2018.

\bibitem{li2018dpi}
Yunzhu Li, Jiajun Wu, Russ Tedrake, Joshua~B Tenenbaum, and Antonio Torralba.
\newblock Learning particle dynamics for manipulating rigid bodies, deformable
  objects, and fluids.
\newblock {\em arXiv preprint arXiv:1810.01566}, 2018.

\bibitem{sanchez2020learning}
Alvaro Sanchez-Gonzalez, Jonathan Godwin, Tobias Pfaff, Rex Ying, Jure
  Leskovec, and Peter~W Battaglia.
\newblock Learning to simulate complex physics with graph networks.
\newblock {\em arXiv preprint arXiv:2002.09405}, 2020.

\bibitem{bruna2013spectral}
Joan Bruna, Wojciech Zaremba, Arthur Szlam, and Yann LeCun.
\newblock Spectral networks and locally connected networks on graphs.
\newblock {\em arXiv preprint arXiv:1312.6203}, 2013.

\bibitem{defferrard2016convolutional}
Micha{\"e}l Defferrard, Xavier Bresson, and Pierre Vandergheynst.
\newblock Convolutional neural networks on graphs with fast localized spectral
  filtering.
\newblock In {\em Advances in neural information processing systems}, pages
  3844--3852, 2016.

\bibitem{henaff2015deep}
Mikael Henaff, Joan Bruna, and Yann LeCun.
\newblock Deep convolutional networks on graph-structured data.
\newblock {\em arXiv preprint arXiv:1506.05163}, 2015.

\bibitem{Ummenhofer2020Lagrangian}
Benjamin Ummenhofer, Lukas Prantl, Nils Thuerey, and Vladlen Koltun.
\newblock Lagrangian fluid simulation with continuous convolutions.
\newblock In {\em International Conference on Learning Representations}, 2020.

\bibitem{sanchez2019hamiltonian}
Alvaro Sanchez-Gonzalez, Victor Bapst, Kyle Cranmer, and Peter Battaglia.
\newblock Hamiltonian graph networks with ode integrators.
\newblock {\em arXiv preprint arXiv:1909.12790}, 2019.

\bibitem{cranmer2020lagrangian}
Miles Cranmer, Sam Greydanus, Stephan Hoyer, Peter Battaglia, David Spergel,
  and Shirley Ho.
\newblock Lagrangian neural networks.
\newblock {\em arXiv preprint arXiv:2003.04630}, 2020.

\bibitem{vaswani2017attention}
Ashish Vaswani, Noam Shazeer, Niki Parmar, Jakob Uszkoreit, Llion Jones,
  Aidan~N Gomez, {\L}ukasz Kaiser, and Illia Polosukhin.
\newblock Attention is all you need.
\newblock In {\em Advances in neural information processing systems}, pages
  5998--6008, 2017.

\bibitem{kipf2016semi}
Thomas~N Kipf and Max Welling.
\newblock Semi-supervised classification with graph convolutional networks.
\newblock {\em arXiv preprint arXiv:1609.02907}, 2016.

\bibitem{Tanabashi:2018oca}
M.~Tanabashi et~al.
\newblock {Review of Particle Physics}.
\newblock {\em Phys. Rev. D}, 98(3):030001, 2018.

\bibitem{Sirunyan:2017ulk}
A.M. Sirunyan et~al.
\newblock {Particle-flow reconstruction and global event description with the
  CMS detector}.
\newblock {\em JINST}, 12(10):P10003, 2017.

\bibitem{henrionneural}
J.~Bruna K.~Cho K.~Cranmer G.~Louppe et~al. I.~Henrion, J.~Brehmer.
\newblock Neural message passing for jet physics.
\newblock In {\em Deep Learning for Physical Sciences Workshop at the 31st
  Conference on Neural Information Processing Systems (NIPS)}, 2017.

\bibitem{Komiske_2019}
Patrick~T. Komiske, Eric~M. Metodiev, and Jesse Thaler.
\newblock Energy flow networks: deep sets for particle jets.
\newblock {\em Journal of High Energy Physics}, 2019(1), Jan 2019.

\bibitem{qu2019particlenet}
Huilin Qu and Loukas Gouskos.
\newblock Particlenet: Jet tagging via particle clouds, 2019.

\bibitem{Moreno_2020}
Eric~A. Moreno, Olmo Cerri, Javier~M. Duarte, Harvey~B. Newman, Thong~Q.
  Nguyen, Avikar Periwal, Maurizio Pierini, Aidana Serikova, Maria Spiropulu,
  and Jean-Roch Vlimant.
\newblock Jedi-net: a jet identification algorithm based on interaction
  networks.
\newblock {\em The European Physical Journal C}, 80(1), Jan 2020.

\bibitem{moreno2019interaction}
Eric~A. Moreno, Thong~Q. Nguyen, Jean-Roch Vlimant, Olmo Cerri, Harvey~B.
  Newman, Avikar Periwal, Maria Spiropulu, Javier~M. Duarte, and Maurizio
  Pierini.
\newblock Interaction networks for the identification of boosted $h\to
  b\overline{b}$ decays, 2019.

\bibitem{mikuni2020abcnet}
Vinicius Mikuni and Florencia Canelli.
\newblock Abcnet: An attention-based method for particle tagging, 2020.

\bibitem{Bernreuther:2020vhm}
Elias Bernreuther, Thorben Finke, Felix Kahlhoefer, Michael Krämer, and
  Alexander Mück.
\newblock {Casting a graph net to catch dark showers}.
\newblock 6 2020.

\bibitem{wang2018dynamic}
Yue Wang, Yongbin Sun, Ziwei Liu, Sanjay~E. Sarma, Michael~M. Bronstein, and
  Justin~M. Solomon.
\newblock Dynamic graph cnn for learning on point clouds, 2018.

\bibitem{Butter:2017cot}
Anja Butter, Gregor Kasieczka, Tilman Plehn, and Michael Russell.
\newblock {Deep-learned Top Tagging with a Lorentz Layer}.
\newblock {\em SciPost Phys.}, 5(3):028, 2018.

\bibitem{ATLAS:2018ibz}
{Performance of mass-decorrelated jet substructure observables for hadronic
  two-body decay tagging in ATLAS}.
\newblock 2018.

\bibitem{NIPS2017_6931}
Manzil Zaheer, Satwik Kottur, Siamak Ravanbakhsh, Barnabas Poczos, Russ~R
  Salakhutdinov, and Alexander~J Smola.
\newblock Deep sets.
\newblock In I.~Guyon, U.~V. Luxburg, S.~Bengio, H.~Wallach, R.~Fergus,
  S.~Vishwanathan, and R.~Garnett, editors, {\em Advances in Neural Information
  Processing Systems 30}, pages 3391--3401. Curran Associates, Inc., 2017.

\bibitem{velickovic2018graph}
Petar Veli{\v{c}}kovi{\'{c}}, Guillem Cucurull, Arantxa Casanova, Adriana
  Romero, Pietro Li{\`{o}}, and Yoshua Bengio.
\newblock {Graph Attention Networks}.
\newblock {\em International Conference on Learning Representations}, 2018.

\bibitem{choma2018graph}
Nicholas Choma, Federico Monti, Lisa Gerhardt, Tomasz Palczewski, Zahra
  Ronaghi, Prabhat, Wahid Bhimji, Michael~M. Bronstein, Spencer~R. Klein, and
  Joan Bruna.
\newblock Graph neural networks for icecube signal classification, 2018.

\bibitem{Abdughani:2018wrw}
Murat Abdughani, Jie Ren, Lei Wu, and Jin~Min Yang.
\newblock {Probing stop pair production at the LHC with graph neural networks}.
\newblock {\em JHEP}, 08:055, 2019.

\bibitem{Ren:2019xhp}
Jie Ren, Lei Wu, and Jin~Min Yang.
\newblock {Unveiling CP property of top-Higgs coupling with graph neural
  networks at the LHC}.
\newblock {\em Phys. Lett. B}, 802:135198, 2020.

\bibitem{Abdughani:2020xfo}
Murat Abdughani, Daohan Wang, Lei Wu, Jin~Min Yang, and Jun Zhao.
\newblock {Probing triple Higgs coupling with machine learning at the LHC}.
\newblock 5 2020.

\bibitem{Bertolini:2014bba}
Daniele Bertolini, Philip Harris, Matthew Low, and Nhan Tran.
\newblock {Pileup Per Particle Identification}.
\newblock {\em JHEP}, 10:059, 2014.

\bibitem{martinez2018pileup}
Jesus~Arjona Martinez, Olmo Cerri, Maurizio Pierini, Maria Spiropulu, and
  Jean-Roch Vlimant.
\newblock Pileup mitigation at the large hadron collider with graph neural
  networks, 2018.

\bibitem{Qasim_2019}
Shah~Rukh Qasim, Jan Kieseler, Yutaro Iiyama, and Maurizio Pierini.
\newblock Learning representations of irregular particle-detector geometry with
  distance-weighted graph networks.
\newblock {\em The European Physical Journal C}, 79(7), Jul 2019.

\bibitem{kieseler2020object}
Jan Kieseler.
\newblock Object condensation: one-stage grid-free multi-object reconstruction
  in physics detectors, graph and image data.
\newblock {\em arXiv preprint arXiv:2002.03605}, 2020.

\bibitem{badiali2020efficiency}
C.~Badiali, F.~A.~Di Bello, G.~Frattari, E.~Gross, V.~Ippolito, M.~Kado, and
  J.~Shlomi.
\newblock Efficiency parameterization with neural networks.
\newblock {\em arXiv preprint arXiv:2004.02665}, 2020.

\bibitem{farrell2018novel}
Steven Farrell, Paolo Calafiura, Mayur Mudigonda, Prabhat, Dustin Anderson,
  Jean-Roch Vlimant, Stephan Zheng, Josh Bendavid, Maria Spiropulu, Giuseppe
  Cerati, Lindsey Gray, Jim Kowalkowski, Panagiotis Spentzouris, and Aristeidis
  Tsaris.
\newblock Novel deep learning methods for track reconstruction, 2018.

\bibitem{Ju:2020xty}
Xiangyang Ju et~al.
\newblock {Graph Neural Networks for Particle Reconstruction in High Energy
  Physics detectors}.
\newblock In {\em {33rd Annual Conference on Neural Information Processing
  Systems}}, 3 2020.

\bibitem{5734880}
W.~{Waltenberger}.
\newblock Rave—a detector-independent toolkit to reconstruct vertices.
\newblock {\em IEEE Transactions on Nuclear Science}, 58(2):434--444, April
  2011.

\bibitem{serviansky2020set2graph}
Hadar Serviansky, Nimrod Segol, Jonathan Shlomi, Kyle Cranmer, Eilam Gross,
  Haggai Maron, and Yaron Lipman.
\newblock Set2graph: Learning graphs from sets, 2020.

\bibitem{johnson2020learning}
Daniel~D Johnson, Hugo Larochelle, and Daniel Tarlow.
\newblock Learning graph structure with a finite-state automaton layer.
\newblock {\em arXiv preprint arXiv:2007.04929}, 2020.

\bibitem{kitaev2020reformer}
Nikita Kitaev, {\L}ukasz Kaiser, and Anselm Levskaya.
\newblock Reformer: The efficient transformer.
\newblock {\em arXiv preprint arXiv:2001.04451}, 2020.

\bibitem{wang2020linformer}
Sinong Wang, Belinda Li, Madian Khabsa, Han Fang, and Hao Ma.
\newblock Linformer: Self-attention with linear complexity.
\newblock {\em arXiv preprint arXiv:2006.04768}, 2020.

\bibitem{li2018learning}
Yujia Li, Oriol Vinyals, Chris Dyer, Razvan Pascanu, and Peter Battaglia.
\newblock Learning deep generative models of graphs.
\newblock {\em arXiv preprint arXiv:1803.03324}, 2018.

\bibitem{nash2020polygen}
Charlie Nash, Yaroslav Ganin, SM~Eslami, and Peter~W Battaglia.
\newblock Polygen: An autoregressive generative model of 3d meshes.
\newblock {\em arXiv preprint arXiv:2002.10880}, 2020.

\bibitem{you2018graphrnn}
Jiaxuan You, Rex Ying, Xiang Ren, William~L Hamilton, and Jure Leskovec.
\newblock Graphrnn: Generating realistic graphs with deep auto-regressive
  models.
\newblock {\em arXiv preprint arXiv:1802.08773}, 2018.

\bibitem{grover2018graphite}
Aditya Grover, Aaron Zweig, and Stefano Ermon.
\newblock Graphite: Iterative generative modeling of graphs.
\newblock {\em arXiv preprint arXiv:1803.10459}, 2018.

\bibitem{wang2018graphgan}
Hongwei Wang, Jia Wang, Jialin Wang, Miao Zhao, Weinan Zhang, Fuzheng Zhang,
  Xing Xie, and Minyi Guo.
\newblock Graphgan: Graph representation learning with generative adversarial
  nets.
\newblock In {\em Thirty-second AAAI conference on artificial intelligence},
  2018.

\bibitem{kipf2018neural}
Thomas Kipf, Ethan Fetaya, Kuan-Chieh Wang, Max Welling, and Richard Zemel.
\newblock Neural relational inference for interacting systems.
\newblock {\em arXiv preprint arXiv:1802.04687}, 2018.

\bibitem{hamrick2018relational}
Jessica~B Hamrick, Kelsey~R Allen, Victor Bapst, Tina Zhu, Kevin~R McKee,
  Joshua~B Tenenbaum, and Peter~W Battaglia.
\newblock Relational inductive bias for physical construction in humans and
  machines.
\newblock {\em arXiv preprint arXiv:1806.01203}, 2018.

\bibitem{bapst2019structured}
Victor Bapst, Alvaro Sanchez-Gonzalez, Carl Doersch, Kimberly~L Stachenfeld,
  Pushmeet Kohli, Peter~W Battaglia, and Jessica~B Hamrick.
\newblock Structured agents for physical construction.
\newblock {\em arXiv preprint arXiv:1904.03177}, 2019.

\bibitem{ying2018hierarchical}
Zhitao Ying, Jiaxuan You, Christopher Morris, Xiang Ren, Will Hamilton, and
  Jure Leskovec.
\newblock Hierarchical graph representation learning with differentiable
  pooling.
\newblock In {\em Advances in neural information processing systems}, pages
  4800--4810, 2018.

\end{thebibliography}
\bibliographystyle{unsrt}

\end{document}